\documentclass[prb, twocolumn, amsmath, amssymb, amsfonts, superscriptaddress]{revtex4}
\usepackage{graphicx}
\usepackage{graphics}
\usepackage{dcolumn}
\usepackage{bm}
\usepackage{amssymb}
\usepackage{amsmath}
\usepackage{amsfonts}
\usepackage{epsfig}

\newcommand {\bra} [1] {\langle #1 |}
\newcommand {\ket} [1] {| #1 \rangle}
\newcommand {\bkt} [1] {\langle #1 \rangle}

\newcommand {\pd} [2] {\frac{\partial #1}{\partial #2}}
\newcommand {\td} [2] {\frac{d #1}{d #2}}

 \newcommand {\beq}{\begin{equation}}
\newcommand {\eeq}{\end{equation}}
\newcommand {\bea}{\begin{eqnarray}}
\newcommand {\eea}{\end{eqnarray}}
\begin{document}
\title{Two-dimensional surface charge transport in topological insulators}
\author{Dimitrie Culcer} 
\affiliation{Condensed Matter Theory Center, Department of Physics, University of Maryland, College Park, Maryland 20742-4111, USA}
\author{E. H. Hwang} 
\affiliation{Condensed Matter Theory Center, Department of Physics, University of Maryland, College Park, Maryland 20742-4111, USA}
\author{Tudor D. Stanescu} 
\affiliation{Department of Physics, West Virginia University, Morgantown WV 26506, USA}
\author{S. Das Sarma}
\affiliation{Condensed Matter Theory Center, Department of Physics, University of Maryland, College Park, Maryland 20742-4111, USA}
\begin{abstract}
We construct a theory of charge transport by the surface states of topological insulators in three dimensions. The focus is on the experimentally relevant case when the Fermi energy $\varepsilon_F$ and transport scattering time $\tau$ satisfy $\varepsilon_F \tau/\hbar \gg 1$, but $\varepsilon_F$ lies below the bottom of the conduction band. Our theory is based on the spin density matrix and takes the quantum Liouville equation as its starting point. The scattering term is determined to linear order in the impurity density $n_i$ and explicitly accounts for the absence of backscattering, while screening is included in the random phase approximation. The main contribution to the conductivity is $\propto n_i^{-1}$ and has different carrier density dependencies for different forms of scattering, while an additional contribution is independent of $n_i$. The dominant scattering angles can be inferred by studying the ratio of the transport time to the Bloch lifetime as a function of the Wigner-Seitz radius $r_s$. The current generates a \textit{spin polarization} that could provide a smoking-gun signature of surface state transport. We also discuss the effect on the surface states of adding metallic contacts.
\end{abstract}
\date{\today}
\maketitle

\section{Introduction}

Topological insulators are a new class of materials that have been intensely researched in recent years. \cite{Hasan_TI_RMP10} They are band insulators in the bulk while conducting along the surfaces, possessing surface states with a spin texture protected by time-reversal symmetry. Within a certain parameter range the surface states are well described by a Dirac cone, allowing for parallels with graphene and relativistic physics, and prohibiting backscattering. Certain physical properties related to the surface states of these materials can be expressed in terms of topologically invariant quantities, and are thus protected against smooth perturbations invariant under time-reversal. For this reason topological insulators are being explored with a view towards applications, as a potential platform for quantum computation,\cite{SDS_TQC_RMP08} and as rich physical systems in their own right.

The concept of a topological insulator dates back to the work of Kane and Mele, who focused on two-dimensional systems. \cite{KaneMele_QSHE_PRL05} These authors demonstrated that, in addition to the well-known TKNN invariant,\cite{TKNN} characterizing the time-reversal breaking quantum Hall state, an additional topological invariant $\nu$ exists, which characterizes the surface states of time-reversal invariant insulators. In two dimensions this invariant can be zero or one, the former representing ordinary insulators (equivalent to the vacuum) and the latter topological insulators. The two band structures corresponding to $\nu = 0$ and $\nu = 1$ cannot be deformed into one another. The invariant $\nu$, which is related to the $Z_2$ topology, counts the number of pairs of Dirac cones on the surfaces. Kane and Mele \cite{KaneMele_QSHE_PRL05} proposed realizing a topological insulator in graphene by using the spin-orbit interaction to open a gap, yet the spin-orbit interaction in graphene is rather small. Subsequently Bernevig \textit{et al.} predicted that a topologically insulating state, the quantum spin Hall insulator, could be realized in HgCdTe quantum wells. \cite{Bernevig_QSHE_Science06} This state was observed shortly after its prediction by K\"onig \textit{et al.} \cite{Koenig_HgTe_QSHE_Science07} (for a review of this effect see Ref.\ \onlinecite{Koenig_QSHE_JPSJ08}.) The quantum spin Hall state constitutes a topological insulator in two dimensions. Soon afterwards topological insulators were predicted to exist in three dimensions, \cite{Fu_3DTI_PRL07, Roy_Z2_PRB09, Moore_TRI_TI_Invariants_PRB07} in which four $Z_2$ topological invariants exist, customarily labeled $\nu_0$, $\nu_1$, $\nu_2$ and $\nu_3$. The $Z_2$ invariants can be calculated easily if the system has inversion symmetry. \cite{Fu_TI_InvSym_PRB07} The invariant $\nu_0$ determines whether a topological insulator is weak ($\nu_0 = 0$) or strong ($\nu_0 = 1$.) The absence of backscattering gives rise to phenomena such as Klein tunneling and is believed to make strong topological insulators immune to Anderson localization. \cite{Hasan_TI_RMP10}

A significant fraction of theoretical research on these materials to date has focused on their topology. A number of articles have been devoted to the classification of topological insulators. \cite{Hatsugai_TI_Char_JPSJ05, Roy_Z2_PRB09, Murakami_SHI_PRL04, Moore_TRI_TI_Invariants_PRB07, Schnyder_Classification_PRB08}  A description of topological insulators in terms of topological field theory has been given in Ref.\ \onlinecite{Qi_TFT_PRB08}, and a general theory of topological insulators and superconductors has been formulated in Ref.\ \onlinecite{Hosur_3DChiralTI_PRB10}.  An effective continuous model for the surface states of three dimensional topological insulators has been proposed in Ref.\ \onlinecite{Shen_3DTI_EffMod_10}. Their effective action in the presence of an electromagnetic field has been studied with a focus on the electric-magnetic duality, \cite{Karch_TI_ElMagDual_PRL09} and analogies have been made with the topological invariants of the Standard Model of particle physics. \cite{Volovik_JETPLett10} The surface states of topological insulators in strong magnetic fields have been investigated in Ref. \onlinecite{Lee_TI_Curved_PRL09}. 

The interface between a topological insulator and a superconductor was predicted to support Majorana fermions, \cite{Fu_Proximity_Majorana_PRL08} that is, particles that are their own antiparticles. Subsequent articles have focused on probing Majorana fermions, \cite{Fu_Majorana_Probe_PRL08} including a scheme employing interferometry. \cite{Akhmerov_TI_MajorInterfere_PRL09} Recently the proximity effect between a superconductor and a topological insulator has been re-examined. \cite{Stanescu_TI_Proximity_10} In addition, other exotic states have also been predicted, including a helical metal, \cite{Ran_TI_HelicalMetal_NP09} superconductivity and ferromagnetism induced by the proximity effect, \cite{Linder_TI_SCFM_10} unconventional superconductivity, \cite{Linder_TI_UnconvSC_PRL10} a quantized anomalous Hall effect in topological insulators doped with magnetic impurities, \cite{Yu_TI_QuantAHE_10} and a new critical state attributed to the Coulomb interaction. \cite{Ostrovsky_TI_InteractCriticality_09} Further expected effects include the possibility of Cooper pair injection at the interface between a topological insulator and a superconductor, \cite{Sato_TI_CooperPairInj_10} the inverse spin-galvanic effect at the interface between a topological insulator and a ferromagnet, \cite{Garate_TI/FM_InverseSpinGalv_09} giant magnetoresistance, \cite{Chu_TI_CohOsc_GMR_PRB09} and a giant Kerr effect. \cite{Tse_TI_MOKE_10}

Variations have also appeared on the theme of topological insulators. Given that superconductors are also gapped, topological superconductors have been predicted to exist. \cite{Hasan_TI_RMP10, Schnyder_Classification_PRB08, Kitaev_Periodic_AIP09, Qi_TSC_PRL09} Developments in different directions include the topological Anderson insulator, \cite{Beenakker_TopAndIns_PRL09, Jiang_HgTe_Numerical_PRB09, Li_TopAndIns_PRL09} the topological Mott insulator,\cite{Raghu_TopMottIns_PRL08} the topological magnetic insulator, \cite{Moore_3DMagIns_TopSfcStt_PRL08} the fractional topological insulator, \cite{Levin_FracTI_PRL09, Lee_TI_EdgeSoliton_PRL07} the topological Kondo insulator, \cite{Dzero_TKI_PRL10} as well as a host of topological phases arising from quadratic, rather than linear, band crossings. \cite{Sun_TI_SponSymBrk_PRL09} Topological phases in transition-metal based strongly correlated systems have been the subject of Refs.\ \onlinecite{Xu_3DTI_TRPhasTrans_PRB10, Pesin_SO_StrongCorr_09}, and a theory relating topological insulators to Mott physics has been proposed in Ref.\ \onlinecite{Rachel_TI_SpinCrgSep_10}.
 
Following the discovery of the quantum spin-Hall state, several materials were predicted to be topological insulators in three dimensions. The first was the alloy Bi$_{1-x}$Sb$_x$,\cite{Teo_BiSb_SfcStt_PRB08, Zhang_BiSb_SfcStt_PRB09} followed by Bi$_2$Se$_3$, Bi$_2$Te$_3$, and Sb$_2$Te$_3$. \cite{Zhang_TI_BandStr_NP09} In particular Bi$_2$Se$_3$ and Bi$_2$Te$_3$ have long been known from thermoelectric transport as displaying sizable Peltier and Seebeck effects, and their high quality has ensured their place at the forefront of experimental attention. \cite{Hasan_TI_RMP10} Initial predictions of the existence of chiral surface states were confirmed by first principles studies of Bi$_2$Se$_3$, Bi$_2$Te$_3$, and Sb$_2$Te$_3$. \cite{Zhang_TI_1stPrinc_10} On an interesting related note, it has recently been proposed that topologically insulating states can be realized in cold atoms, \cite{Stanescu_AMO_TI_PRA09} though these lie beyond the scope of the present work.

After a timid start, experimental progress on the materials above has skyrocketed. Hsieh \textit{et al}, \cite{Hsieh_BiSb_QSHI_Nature08} using angle-resolved photoemission spectroscopy (ARPES), were the first to observe the Dirac cone-like surface states of Bi$_{1-x}$Sb$_x$, with $x\approx 0.1$. Following that, the same group used spin-resolved ARPES to measure the spin polarization of the surface states, demonstrating the correlation between spin and momentum.\cite{Hsieh_BiSb_QmSpinTxtr_Science09} Shortly afterwards experiments identified the Dirac cones of Bi$_2$Se$_3$,\cite{Xia_Bi2Se3_LargeGap_NP09} Bi$_2$Te$_3$,\cite{Chen_Bi2Te3_Science09, Hsieh_Bi2Te3_Sb2Te3_PRL09} and Sb$_2$Te$_3$.\cite{Hsieh_Bi2Te3_Sb2Te3_PRL09} Recently Bi$_2$Se$_3$ nanowires and nanoribbons have also been grown \cite{Kong_TI_WireRbn_NanoLett10} and experiments have begun to investigate the transition between the three- and two-dimensional phases in this material.\cite{Zhang_Bi2Se3_Crossover_09} Bi$_2$Se$_3$ has also been studied in a magnetic field, revealing a magnetoelectric coupling \cite{LaForge_BisSe3_MagnetoEl_09} and an oscillatory angular dependence of the magnetoresistance. \cite{Taskin_BiSb_MagRes_10} Sample growth dynamics of Bi$_2$Te$_3$ by means of MBE were monitored by reflection high-energy electron diffraction (RHEED). \cite{Li_Bi2Se3_GrowthDyn_09} Doping Bi$_2$Te$_3$ with Mn has been shown to result in the onset of ferromagnetism, \cite{Hor_DopedTI_FM_10} while doping Bi$_2$Se$_3$ with Cu leads to superconductivity.\cite{Hor_CuBi2Se3_SC_PRL10, Wray_CuBi2Se3_SC_09}

Considerable efforts have been devoted to scanning tunneling microscopy (STM) and spectroscopy (STS), which enable the study of quasiparticle scattering. Scattering off surface defects, in which the initial state interferes with the final scattered state, results in a standing-wave interference pattern with a spatial modulation determined by the momentum transfer during scattering. These manifest themselves as oscillations of the local density of states in real space, which have been seen in several materials with topologically protected surface states. Roushan \textit{et al.} \cite{Roushan_BiSb_STM_NoBxct_Nat09} have imaged the surface states of Bi$_{0.92}$Sb$_{0.08}$ using STM and spin-resolved ARPES demonstrating that, as expected theoretically, in the absence of spin-independent (time-reversal invariant) disorder, scattering between states of opposite momenta is strongly suppressed. Gomes \textit{et al.} \cite{Gomes_Sb_STM_NoBxct_09} investigated the (111) surface of Sb, which displays a topological metal phase, and found a similar suppression of backscattering. Zhang \textit{et al.} \cite{Zhang_Bi2Te3_STM_NoBxct_PRL09} and Alpichshev \textit{et al.}\cite{Alp_Bi2Te3_STM_NoBxct_PRL10} have observed the suppression of backscattering in Bi$_2$Te$_3$. Theories have recently been formulated to describe the quasiparticle interference seen in STM experiments, accounting for the absence of backscattering. \cite{Guo_TI_QptIntfrnc_PRB10, Lee_TI_Qpt_Intfrnc_PRB09, Park_Bi2Se3_QPtSct_09} Related developments have seen the imaging of a surface bound state using scanning tunneling microscopy. \cite{Alpichshev_Bi2Se3_BoundStt_10} These experimental studies provide evidence of time-reversal symmetry protection of the chiral surface states. 

Despite the success of photoemission and scanning tunneling spectroscopy in identifying chiral surface states, signatures of the surface Dirac cone have not yet been observed in transport. Simply put, the band structure of topological insulators can be visualized as a band insulator with a Dirac cone within the bulk gap, and to access this cone one needs to ensure the chemical potential lies below the bottom of the conduction band. Given that the static dielectric constants of materials under investigation are extremely large, approximately 100 in Bi$_2$Se$_3$ and 200 in Bi$_2$Te$_3$, gating in order to bring the chemical potential down is challenging. Therefore, in all materials studied to date, residual conduction from the bulk exists due to unintentional doping. One potential way forward was opened by Hor \textit{et al.}, \cite{Hor_Bi2Se3_Ca_p_PRB09} who demonstrated that Ca doping ($\approx 1 \%$ substituting for Bi) brings the Fermi energy into the valence band, followed by the demonstration by Hsieh \textit{et al.} \cite{Hsieh_BiSb_QmSpinTxtr_Science09} that Ca doping can be used to tune the Fermi energy so that it lies in the gap. Checkelsky \textit{et al.} \cite{Checkelsky_Bi2Se3_QmIntfr_PRL09} performed transport experiments on Ca-doped samples noting an increase in the resistivity, but concluded that surface state conduction alone could not be responsible for the smallness of the resistivities observed at low temperature. Eto \textit{et al.} \cite{Eto_Bi2Se3_SdH_3DFS_10} reported Shubnikov-de Haas (SdH) oscillations due only to the 3D Fermi surface in Bi$_2$Se$_3$ with an electron density of $5 \times 10^{18}$cm$^{-3}$, while Analytis \textit{et al.} \cite{Analytis_Bi2Se3_ARPES_SdH_3DFS_10} used ARPES and SdH oscillations to probe Bi$_2$Se$_3$. The two methods were found to agree for bulk carriers, but while ARPES identified Dirac cone-like surface states, even at number densities of the order of $10^{17}$cm$^{-3}$ SdH oscillations identified only a bulk Fermi surface, and the surface states were not seen in transport. A series of recent transport experiments claim to have observed separate signatures of bulk and surface conduction. Checkelsky \textit{et al.} \cite{Checkelsky_Bi2Se3_SfcSttCond_10} performed weak localization and universal conductance fluctuation measurements, claiming to have observed surface conduction, although the chemical potential still lay in the conduction band. Steinberg \textit{et al.} \cite{Steinberg_Bi2Se3_Ambipolar_10} reported signatures of ambipolar transport analogous to graphene. However, in this work also, two contributions to transport were found, the bulk contribution was subtracted so as to obtain the surface contribution, and the surface effect was observed only for top gate voltage sweeps, not back gate sweeps. Butch \textit{et al.} \cite{Butch_Bi2Se3_Search_10} carried out SdH, ordinary Hall effect and reflectivity measurements and concluded that the observed signals could be ascribed solely to bulk states. No experimental group has reported SdH oscillations due to the surface states, which are expected at a very different frequency. \cite{Butch_Bi2Se3_Search_10} The above presentation makes it plain that, despite the presence of the Dirac cone, currently no experimental group has produced a true topological insulator. Consequently, all existing topological insulator/metal systems are in practice either heavily bulk-doped materials or thin films of a few monolayers\cite{Zhang_Bi2Se3_Crossover_09} (thus, by definition, not 3D.)

In an intriguing twist, two recent works have reported STM imaging of the Landau levels of Bi$_2$Se$_3$. \cite{Cheng_TI_STM_LL_10, Hanaguri_TI_STM_LL_10} Cheng \textit{et al.}\cite{Cheng_TI_STM_LL_10} claim to have observed Landau levels $n$ analogous to graphene, with energies $\propto{\sqrt{nB}}$ but the expected spin and valley degeneracy of one, yet puzzling features such as the absence of Landau levels \textit{below} the Fermi energy and enhanced oscillations near the conduction band edge have not been resolved, and some arbitrariness is involved in extracting the Landau level filling factor $\nu$. Hanaguri \textit{et al.}\cite{Hanaguri_TI_STM_LL_10} report similar findings in the same material, with analogous features in the Landau level spectrum. Following the pattern set by the imaging of the surface Dirac cone, the Landau levels observed in STM studies have not been seen in transport. These last works illustrate both the enormous potential of the field and the challenges to be overcome experimentally.

The focus of experimental research has shifted to the separation of bulk conduction from surface conduction, hence it is necessary to know what type of contributions to expect from the surface conduction. The only transport theories we are aware of have been Ref.\ \onlinecite{Mondal_TI_Magneto_10}, which focused on magnetotransport in the impurity-free limit, and Ref.\ \onlinecite{Burkov_TI_SpinCharge_10}, which focused on spin-charge coupling. A comprehensive theory of transport in topological insulators has not been formulated to date. This article seeks to remedy the situation. We begin with the quantum Liouville equation, deriving a kinetic equation for the density matrix of topological insulators including the full scattering term to linear order in the impurity density. Peculiar properties of topological insulators, such as the absence of backscattering (which leads to Klein tunneling and other phenomena), are built into our theory. We determine the polarization function of topological insulators. By solving the kinetic equation we determine all contributions to the conductivity to orders one and zero in the transport scattering time $\tau$. We find that the scalar part of the Hamiltonian makes an important contribution to the conductivity, which has a different carrier density dependence than the spin-dependent part, and may be a factor in distinguishing the surface conduction from the bulk conduction. We consider charged impurity as well as short-range surface roughness scattering and determine the density-dependence of the conductivity when either of these is dominant. Most importantly, we find that the electrical current generates a \textit{spin polarization} in the in-plane direction perpendicular to the field, providing a unique signature of surface transport. Moreover, the ratio of the single-particle scattering time to the transport scattering time provides important information on the dominant angles in scattering, which in turn provide information on the dominant scattering mechanisms. At zero temperature, which constitutes the focus of our work, we expect a competition between charged impurity scattering and short-range scattering. For each of these two scattering mechanisms the ratio of the single-particle scattering time to the transport scattering time has a qualitatively different dependence on the Wigner-Seitz radius $r_s$, which parametrizes the relative strength of the kinetic energy and electron-electron interactions. We study these aspects of the transport problem in detail, covering the possible situations that can be encountered experimentally with currently available samples and technology. Finally, we consider the effect on the topological surface states of adding metal contacts, as would be the case in experiment.

Our work shows that the suppression of backscattering seen in experimental work indicates protection against localization resulting from backscattering, rather than protection against resistive scattering in general. The conductivity due to the surface states is determined by scattering of carriers by disorder and defects, the density of which is in turn determined by sample quality. Strong scattering will lead to low mobility, and the mobility is not a topologically protected quantity. To see surface transport one therefore requires very clean samples.

Several features set topological insulators apart from graphene: the twofold valley degeneracy of graphene is not present in topological insulators, while the Hamiltonian is a function of the real spin, rather than pseudospin. This implies that spin dynamics will be qualitatively very different from graphene. Furthermore, the Dirac Hamiltonian is in fact a Rashba Hamiltonian expressed in terms of a rotated spin. Despite the apparent similarities, the study of topological insulators is thus not a simple matter of translating results known from graphene. Due to the dominant spin-orbit interaction it is also very different from ordinary spin-orbit  semiconductors.

The outline of this paper is as follows. In Sec. \ref{sec:Ham} we introduce the effective Hamiltonian for the surface states of topological insulators and briefly discuss its structure. Next, in Sec. \ref{sec:kineq} we derive a kinetic equation for the spin density matrix of topological insulators, including the full scattering term in the presence of an arbitrary elastic scattering potential to linear order in the impurity density. This equation is solved in Sec. \ref{sec:sol}, yielding the contributions to orders one and zero in the transport scattering time $\tau$, and used to determine the charge current. Sec. \ref{sec:expt} discusses our results in the context of experiments, focusing on the electron density and impurity density dependence of the conductivity, ways to determine the dominant angles in impurity scattering, and the effect of the contacts on the surface states. We end with a summary and conclusions.

\section{Effective Hamiltonian for topological insulators} 
\label{sec:Ham}

The effective Hamiltonian describing the surface states of topological insulators can be written in the form
\begin{equation}\label{Ham}
H_{0{\bm k}} = D \, k^2 + A \, {\bm \sigma} \cdot {\bm k}.
\end{equation}
It is understood that ${\bm k} = (k_x, k_y)$. Near ${\bm k} = 0$, when the spectrum is accurately described by a Dirac cone, the scalar term is overwhelmed by the spin-dependent term and can be safely neglected, the remaining Hamiltonian being similar to that of graphene, with the exception that ${\bm \sigma}$ represents the true (rotated) electron spin. On the other hand, at finite doping and at usual electron densities of $10^{11}-10^{12}$cm$^{-2}$ the scalar term is no longer a small perturbation, and should be taken into account. The scalar and spin-dependent terms in the Hamiltonian are of the same magnitude when $D \, k_F = A$, corresponding to $k_F \approx 10^9$ m$^{-1}$ in Bi$_2$Se$_3$, that is, a density of $10^{14}$cm$^{-2}$, which is higher than the realistic densities in transport experiments. Consequently, in this article the terms $\propto D$ will be treated as a perturbation. The eigenenergies are denoted by $\epsilon_\lambda = Dk^2 + \lambda Ak$, with $\lambda = \pm$. We do not take into account small anisotropy terms in the scalar part of the Hamiltonian, such as those discussed in Ref. \onlinecite{Zhang_TI_BandStr_NP09}. Terms cubic in ${\bm k}$ in the spin-orbit interaction are in principle also present, in turn making the energy dispersion anisotropic, but they are much smaller than the ${\bm k}$-linear and quadratic terms in Eq.\ (\ref{Ham}) and are not expected to contribute significantly to quantities discussed in this work.

There are contributions to the charge current from the scalar and spin parts of the density matrix. The current operator is given by
\begin{equation}\label{j}
{\bm j} = -\frac{e}{\hbar} \,( 2D \, {\bm k} + A \, {\bm \sigma}).
\end{equation}
Unlike graphene, where the effective Dirac Hamiltonian is valid even at large electron densities and the current operator is independent of ${\bm k}$ for all ranges of wave vector relevant to transport, in topological insulators the scalar term in the Hamiltonian quadratic in $k$ also contributes to the current, and will yield a different number-density dependence. 

Transport experiments on topological insulators seek to distinguish the conduction due to the surface states from that due to the bulk states. As mentioned in the introduction, it has proven challenging to lower the chemical potential beneath the bottom of the conduction band, so that the materials studied at present are not strictly speaking insulators. It is necessary for $\varepsilon_F$ to be below the bulk conduction band, so that one can be certain that there is only surface conduction, and at the same time the requirement that $\varepsilon_F \tau/\hbar \gg 1$ is necessary for the kinetic equation formalism to be applicable. These assumptions will be made in our work.

\section{Kinetic equation}
\label{sec:kineq}

We wish to derive a kinetic equation for the system driven by an electric field in the presence of random, uncorrelated impurities. To this end we begin with the quantum Liouville equation for the density operator $\hat \rho$,
\begin{equation}
\td{\hat\rho}{t} + \frac{i}{\hbar} \, [\hat{H}, \hat \rho] = 0,
\end{equation}
where the full Hamiltonian $\hat{H} = \hat{H}_0 + \hat{H}_E + \hat U$, the band Hamiltonian $\hat{H}_0$ is defined in Eq.\ (\ref{Ham}), $\hat{H}_E = e{\bm E}\cdot\hat{\bm r}$ represents the interaction with external fields, $\hat{\bm r}$ is the position operator, and $\hat{U}$ is the impurity potential. We consider a set of time-independent states $\{ \ket{{\bm k}s} \}$, where ${\bm k}$ indexes the wave vector and $s$ the spin. The matrix elements of $\hat \rho$ are $\rho_{{\bm k}{\bm k}'} \equiv \rho^{ss'}_{{\bm k}{\bm k}'} = \bra{{\bm k}s} \hat\rho \ket{{\bm k}'s'}$, with the understanding that $\rho_{{\bm k}{\bm k}'}$ is a matrix in spin space. The terms $\hat{H}_0$ and $\hat{H}_E$ are diagonal in wave vector but off-diagonal in spin, while for elastic scattering in the first Born approximation $U_{{\bm k}{\bm k}'}^{ss'} = U_{{\bm k}{\bm k}'}\delta_{ss'}$. The impurities are assumed uncorrelated and the average of $\bra{{\bm k}s}\hat U\ket{{\bm k}'s'}\bra{{\bm k}'s'}\hat U\ket{{\bm k}s}$ over impurity configurations is $(n_i |\bar{U}_{{\bm k}{\bm k}'}|^2 \delta_{ss'})/V$, where $n_i$ is the impurity density, $V$ the crystal volume and $\bar{U}_{{\bm k}{\bm k}'}$ the matrix element of the potential of a single impurity.

The density matrix $\rho_{{\bm k}{\bm k}'}$ is written as $\rho_{{\bm k}{\bm k}'} = f_{{\bm k}} \, \delta_{{\bm k}{\bm k}'} + g_{{\bm k}{\bm k}'}$, where $f_{\bm k}$ is diagonal in wave vector (i.e. $f_{{\bm k}{\bm k}'} \propto \delta_{{\bm k}{\bm k}'}$) while $g_{{\bm k}{\bm k}'}$ is off-diagonal in wave vector (i.e. ${\bm k} \ne {\bm k}'$ always in $g_{{\bm k}{\bm k}'}$.) The quantity of interest in determining the charge current is $f_{\bm k}$, since the current operator is diagonal in wave vector. We therefore derive an effective equation for this quantity by first breaking down the quantum Liouville equation into
\begin{subequations}
\begin{eqnarray}
\label{eq:f} \td{f_{{\bm k}}}{t} + \frac{i}{\hbar} \, [H_{0{\bm k}}, f_{{\bm k}}] & = & - \frac{i}{\hbar} \, [H^E_{\bm k}, f_{{\bm k}}] - \frac{i}{\hbar} \, [\hat U, \hat g]_{{\bm k}{\bm k}} , \\ [1ex]
\label{eq:g} \td{g_{{\bm k}{\bm k}'}}{t} + \frac{i}{\hbar} \, [\hat{H}, \hat g]_{{\bm k}{\bm k}'} & = & - \frac{i}{\hbar} \, [\hat U, \hat f + \hat g]_{{\bm k}{\bm k}'},
\end{eqnarray}
\end{subequations}
and solving Eq.\ (\ref{eq:g}) to first order in $\hat{U}$
\begin{equation}\label{eq:gsol}
g_{{\bm k}{\bm k}'} = - \frac{i}{\hbar} \, \int_0^\infty dt'\, e^{- i \hat H t'/\hbar} \left[\hat U, \hat f (t - t') \right] e^{i \hat H t'/\hbar}\bigg|_{{\bm k}{\bm k}'}.
\end{equation}
The details of this derivation have been included in Appendix \ref{sec:app}. We are interested in variations which are slow on the scale of the momentum relaxation time, consequently we do not take into account memory effects and $\hat f(t - t') \approx \hat{f} (t)$, so that $f_{\bm k}$ obeys
\begin{subequations}\label{eq:FermiJfk}
\begin{eqnarray}
\td{f_{{\bm k}}}{t} & + & \frac{i}{\hbar} \, [H_{0{\bm k}}, f_{{\bm k}}] + \hat J (f_{{\bm k}}) = - \frac{i}{\hbar} \, [H^E_{\bm k}, f_{{\bm k}}], \\ [1ex]
\label{eq:Jfk} \hat J (f_{{\bm k}}) & = & \frac{1}{\hbar^2} \! \int_{0}^\infty \!\!\! dt'\, \left[\hat U, e^{- i \hat H t'/\hbar} \left[\hat U, \hat f (t) \right] e^{ i \hat H t'/\hbar} \right]_{{\bm k}{\bm k}}.
\end{eqnarray}
\end{subequations}
The integral in Eq.\ (\ref{eq:Jfk}), which represents the scattering term, is performed by inserting a regularizing factor $e^{- \eta t'}$ and letting $\eta \rightarrow 0$ in the end. We consider a potential $|U_{{\bm k}{\bm k}'}| \propto \openone$, which does not have spin dependence, and carry out an average over impurity configurations, keeping the terms to linear order in the impurity density $n_i$. The scattering term has the form
\begin{widetext}
\begin{equation}\label{eq:Jfinal} \hat J(f_{\bm k}) = \frac{n_i}{\hbar^2} \lim_{\eta \rightarrow 0} \int \frac{d^2k'}{(2\pi)^2} \, |\bar{U}_{{\bm k}{\bm k}'}|^2
\int^{\infty}_0 dt'\, e^{- \eta t'} \Bigl\{ e^{- i H_{{\bm k}'} t'/\hbar}\big(f_{\bm k} - f_{\bm k}' \big) \,
e^{i H_{{\bm k}} t'/\hbar} +  h. c. \Bigr\}.
\end{equation}
\end{widetext}
For this term to be cast in a useful form one needs to complete the time integral and analyze the spin structure of the scattering operator $\hat J(f_{\bm k})$, which is done below.

\subsection{Scattering term}

The ${\bm k}$-diagonal part of the density matrix $f_{\bm k}$ is a $2 \times 2$ Hermitian matrix, which is decomposed into a scalar part and a spin-dependent part. We write $f_{\bm k} = n_{\bm k}\openone + S_{\bm k}$, where $n_{\bm k}$ represents the scalar part and $\openone$ is the identity matrix in two dimensions. The component $S_{\bm k}$ is itself a $2 \times 2$ Hermitian matrix, which represents the spin-dependent part of the density matrix and is written purely in terms of the Pauli $\sigma$ matrices. Thus, the matrix $S_{\bm k}$ can be expressed as $S_{\bm k} = \frac{1}{2}\, {\bm S}_{{\bm k}} \cdot {\bm \sigma} \equiv \frac{1}{2}\, S_{{\bm k}i} \sigma_i$. With this notation, the scattering term is in turn decomposed into scalar and spin-dependent parts
\begin{widetext}
\begin{equation}
\arraycolsep 0.3ex
\begin{array}{rl}
\displaystyle \hat J(f) = & \displaystyle \frac{n_i}{\hbar^2} \int \frac{d^2k'}{(2\pi)^2} \, |\bar{U}_{{\bm k}{\bm k}'}|^2 \big(n_{\bm k} -
n_{{\bm k}'} \big) \int^{\infty}_0 dt'\, e^{- \eta t'} e^{- i H_{{\bm k}'} t'/\hbar} \, e^{i H_{{\bm k}} t'/\hbar} + h.c. \\ [3ex] 
+ & \displaystyle \frac{n_i}{2\hbar^2} \int \frac{d^2k'}{(2\pi)^2} \, |\bar{U}_{{\bm k}{\bm k}'}|^2 \big( {\bm S}_{\bm k} - {\bm S}_{{\bm k}'} \big) \cdot
\int^{\infty}_0 dt'\, e^{-\eta t'} e^{- i H_{{\bm k}'} t'/\hbar} {\bm \sigma}  \, e^{i H_{{\bm k}} t'/\hbar} + h.c. 
\end{array}
\end{equation}
We assume electron doping, which implies that the chemical potential is in the conduction band $\epsilon_+$. For the scalar part $n_{\bm k}$ and spin-dependent part $S_{\bm k}$ of the density matrix the above results in
\begin{equation}
\arraycolsep 0.3ex
\begin{array}{rl}
\displaystyle e^{- i\hat H_{{\bm k}'} t'/\hbar} e^{ i \hat H_{{\bm k}}t'/\hbar} + h.c. \rightarrow & \displaystyle \frac{\pi\hbar}{2}\, (1 + \hat{\bm k}\cdot\hat{\bm k}') \, [\delta(\epsilon'_+ - \epsilon_+) + \delta(\epsilon'_- - \epsilon_-) ]  \\ [3ex]
+ & \displaystyle \frac{\pi\hbar}{2}\, {\bm \sigma}\cdot(\hat{\bm k} + \hat{\bm k}') \, [\delta(\epsilon'_+ - \epsilon_+) - \delta(\epsilon'_- - \epsilon_-) ] \\ [3ex]
\displaystyle e^{- i\hat H_{{\bm k}'} t'/\hbar}{\bm \sigma}\, e^{ i \hat H_{{\bm k}}t'/\hbar} + h. c. \rightarrow & \displaystyle \frac{\pi\hbar}{2} \, [{\bm \sigma}(1 - \hat{\bm k} \cdot \hat{\bm k}') + (\hat{\bm k} \cdot{\bm \sigma}) \, \hat{\bm k}' + \hat{\bm k} \, (\hat{\bm k}'\cdot{\bm \sigma})] \, [\delta(\epsilon'_+ - \epsilon_+) + \delta(\epsilon'_- - \epsilon_-)] \\ [3ex]
+ & \displaystyle \frac{\pi\hbar}{2} \, (\hat{\bm k} + \hat{\bm k}') \, [\delta(\epsilon'_+ - \epsilon_+) - \delta(\epsilon'_- - \epsilon_-)], 
\end{array}
\end{equation}
where $\hat{\bm k}$ is a unit vector in the direction of ${\bm k}$. We emphasize that there is no backscattering, so the physics of topological insulators is built into the kinetic equation from the beginning. Collecting all terms, we obtain the scattering term as $\hat J(f) = \hat J(n) + \hat J(S)$, with
\begin{equation}
\arraycolsep 0.3ex
\begin{array}{rl}
\displaystyle \hat J(n) = & \displaystyle \frac{\pi n_i}{2\hbar} \int \frac{d^2k'}{(2\pi)^2} \, |\bar{U}_{{\bm k}{\bm k}'}|^2 \big(n_{\bm k} -
n_{{\bm k}'} \big) (1 + \hat{\bm k}\cdot\hat{\bm k}') \, [\delta(\epsilon'_+ - \epsilon_+) + \delta(\epsilon'_- - \epsilon_-) ] \\ [3ex]
+ & \displaystyle \frac{\pi n_i}{2\hbar} \int \frac{d^2k'}{(2\pi)^2} \, |\bar{U}_{{\bm k}{\bm k}'}|^2 \big(n_{\bm k} -
n_{{\bm k}'} \big)  {\bm \sigma}\cdot(\hat{\bm k} + \hat{\bm k}')\, [\delta(\epsilon'_+ - \epsilon_+) - \delta(\epsilon'_- - \epsilon_-) ] \\ [3ex]
\displaystyle \hat J(S) = & \displaystyle \frac{\pi n_i}{4\hbar} \int \frac{d^2k'}{(2\pi)^2} \, |\bar{U}_{{\bm k}{\bm k}'}|^2 \big( {\bm S}_{{\bm k}} - {\bm S}_{{\bm k}'}\big)\cdot [{\bm \sigma}(1 - \hat{\bm k} \cdot \hat{\bm k}') + (\hat{\bm k} \cdot{\bm \sigma}) \, \hat{\bm k}' + \hat{\bm k} \, (\hat{\bm k}'\cdot{\bm \sigma}) ] \, [\delta(\epsilon'_+ - \epsilon_+) + \delta(\epsilon'_- - \epsilon_-) ] \\ [3ex]
+ & \displaystyle \frac{\pi n_i}{4\hbar} \int \frac{d^2k'}{(2\pi)^2} \, |\bar{U}_{{\bm k}{\bm k}'}|^2 \big( {\bm S}_{{\bm k}} - {\bm S}_{{\bm k}'}\big)\cdot (\hat{\bm k} + \hat{\bm k}') \,[\delta(\epsilon'_+ - \epsilon_+) - \delta(\epsilon'_- - \epsilon_-) ] .
\end{array}
\end{equation}
The details of this calculation are contained in Appendix \ref{sec:app}. The energy $\delta$-functions $\delta(\epsilon'_\pm - \epsilon_\pm) = [1/(A \pm 2Dk)] \, \delta(k' - k)$. The $\delta$-functions of $\epsilon_-$ are needed in the scattering term in order to ensure agreement with Boltzmann transport. This fact is already seen for a Dirac cone dispersion in graphene where the expression for the conductivity found in Ref.\ \onlinecite{Culcer_Gfn_Transp_PRB08} using the density-matrix formalism agrees with the Boltzmann transport formula of Ref.\ \onlinecite{SDS_Gfn_RMP10} (the definition of $\tau$ differs by a factor of two in these two references.) The necessity of keeping the $\epsilon_-$ terms is a result of Zitterbewegung: the two branches $\lambda = \pm 1$ are mixed, thus scattering of an electron requires conservation of $\epsilon_+$ as well as $\epsilon_-$. In the equivalent kinetic equation for graphene there is no coupling of the scalar and spin distributions because of the particle-hole symmetry inherent in the Dirac Hamiltonian. In topological insulators the $Dk^2$ term breaks particle-hole symmetry and gives rise to a small coupling of the scalar and spin distributions. Later below we quantify this coupling by an effective time $\tau_D$, which $\rightarrow \infty$ as $D \rightarrow 0$.

We find a series of scattering terms coupling the scalar and spin distributions. Let $\gamma$ represent the angle between $\hat{\bm k}$ and $\hat{\bm k}'$, and the unit vector $\hat{\bm \theta}$ in polar coordinates represent the direction perpendicular to $\hat{\bm k}$, that is the tangential direction. The unit vectors $\hat{\bm k}$, $\hat{\bm k}'$, $\hat{\bm \theta}$ and $\hat{\bm \theta}'$ are related by the transformations
\begin{equation}
\arraycolsep 0.3ex
\begin{array}{rl}
\displaystyle \hat{\bm k}' = & \displaystyle \hat{\bm k} \cos\gamma
+ \hat{\bm \theta}\sin\gamma \\ [3ex] 
\displaystyle \hat{\bm \theta}' = & \displaystyle - \hat{\bm k} \sin\gamma
+ \hat{\bm \theta}\cos\gamma \\ [3ex] 
\displaystyle \hat{\bm k} = &
\displaystyle \hat{\bm k}' \cos\gamma - \hat{\bm \theta}'
\sin\gamma.
\end{array}
\end{equation}
We decompose the matrix $S_{\bm k} = S_{{\bm k}\parallel} + S_{{\bm k}\perp}$ and write those two parts in turn as $S_{{\bm k}\parallel} = (1/2) \, s_{{\bm k}\parallel} \, \sigma_{{\bm k} \parallel}$ and $S_{{\bm k}\perp} = (1/2) \, s_{{\bm k}\perp} \sigma_{{\bm k}\perp}$. The small $s_{{\bm k}\parallel}$ and $s_{{\bm k}\perp}$ are scalars and are given by $s_{{\bm k}\parallel} = {\bm S}_{\bm k} \cdot\hat{\bm k}$ and $s_{{\bm k}\perp} = {\bm S}_{\bm k} \cdot\hat{\bm \theta}$, and similarly $\sigma_{{\bm k} \parallel} = {\bm \sigma}\cdot\hat{\bm k}$ and $\sigma_{{\bm k}\perp} = {\bm \sigma}\cdot\hat{\bm \theta}$. We introduce projection operators $P_n$, $P_\parallel$ and $P_\perp$ onto the scalar part, $\sigma_{{\bm k} \parallel}$ and $\sigma_{{\bm k} \perp}$ respectively. We will need the following projections of the scattering term acting on the spin-dependent part of the density matrix
\begin{equation}
\arraycolsep 0.3ex
\begin{array}{rl}
\displaystyle P_\parallel \hat J(S_{{\bm k}\parallel}) = & \displaystyle \frac{kn_i}{16\hbar \pi} \, \bigg[ \frac{1}{(A + 2Dk)} + \frac{1}{(A - 2Dk)} \bigg] \int d\theta' \, |\bar{U}_{{\bm k}{\bm k}'}|^2 \, (s_{{\bm k}\parallel} - s_{{\bm k}'\parallel})(1 + \cos\gamma)\, \sigma_{{\bm k}\parallel} \\ [3ex]
\displaystyle P_\perp \hat J(S_{{\bm k}\parallel}) = & \displaystyle \frac{kn_i}{16\hbar \pi} \, \bigg[ \frac{1}{(A + 2Dk)} + \frac{1}{(A - 2Dk)} \bigg] \int\! d\theta' \, |\bar{U}_{{\bm k}{\bm k}'}|^2\, (s_{{\bm k}\parallel} - s_{{\bm k}'\parallel}) \sin\gamma \, \sigma_{{\bm k} \perp} \\ [3ex]
\displaystyle P_\parallel \hat J(S_{{\bm k}\perp}) = & \displaystyle \frac{kn_i}{16\hbar \pi} \, \bigg[ \frac{1}{(A + 2Dk)} + \frac{1}{(A - 2Dk)} \bigg] \int\! d\theta' \, |\bar{U}_{{\bm k}{\bm k}'}|^2\, \big(s_{{\bm k}\perp} + s_{{\bm
k}'\perp}\big) \sin\gamma \sigma_{{\bm k}\parallel}.
\end{array}
\end{equation}
and $P_\perp \hat J(S_{{\bm k}\perp})$ will be deferred to a later time. Furthermore, the scattering terms coupling the scalar and spin distributions are
\begin{equation}
\arraycolsep 0.3ex
\begin{array}{rl}
\displaystyle \hat J_{n \rightarrow S} (n_{\bm k}) = & \displaystyle \frac{\pi n_i}{2\hbar} \int \frac{d^2k'}{(2\pi)^2} \, |\bar{U}_{{\bm k}{\bm k}'}|^2 \big(n_{\bm k} -
n_{{\bm k}'} \big) [\sigma_{{\bm k}\parallel} (1 + \cos\gamma) + \sigma_{{\bm k}\perp} \sin\gamma]\, [\delta(\epsilon'_+ - \epsilon_+) - \delta(\epsilon'_- - \epsilon_-) ] \\ [3ex]
\displaystyle \hat J_{S \rightarrow n} (S_{\bm k}) = & \displaystyle \frac{\pi n_i}{4\hbar} \int \frac{d^2k'}{(2\pi)^2} \, |\bar{U}_{{\bm k}{\bm k}'}|^2 \big[ (s_{{\bm k}\parallel} - s_{{\bm k}'\parallel})(1 + \cos\gamma) + (s_{{\bm k}\perp} + s_{{\bm k}'\perp})\sin\gamma \big] \, [\delta(\epsilon'_+ - \epsilon_+) - \delta(\epsilon'_- - \epsilon_-) ].
\end{array}
\end{equation}
\end{widetext}
Notice the factors of $(1 + \cos\gamma)$ which prohibit backscattering (and lead to Klein tunneling.)

\subsection{Polarization function and effective scattering potential}

The above form of the scattering term is valid for any scattering potential, as long as scattering is elastic. To make our analysis more specific we consider a screened Coulomb potential, and evaluate the screening function in the random phase approximation. In this approximation the polarization function is obtained by summing the lowest bubble diagram, and takes the form \cite{Hwang_Gfn_Screening_PRB07}
\begin{equation}
\Pi (q, \omega) = -\frac{1}{A} \sum_{{\bm k}\lambda\lambda'} \frac{f_{0{\bm k}\lambda} - f_{0{\bm k}'\lambda'}}{\omega + \varepsilon_{{\bm k}\lambda} - \varepsilon_{{\bm k}'\lambda'} + i\eta} \, \bigg(\frac{1 + \lambda\lambda' \cos\gamma}{2}\bigg), 
\end{equation}
where $f_{0{\bm k}\lambda} \equiv f_0(\varepsilon_{{\bm k}\lambda})$ is the equilibrium Fermi distribution function. The static dielectric function, of interest to us in this work, can be written as $\epsilon(q) = 1 + v(q) \,\Pi(q)$, where $v(q) = e^2/(2\epsilon_0\epsilon_r q)$. To determine $\Pi (q)$, we assume $T=0$ and use the Dirac cone approximation as in Ref.\ \onlinecite{Hwang_Gfn_Screening_PRB07}. This approximation is justified by the fact that we are working in a regime in which $T/T_F \ll 1$, with $T_F$ the Fermi temperature. The results obtained are in fact correct to linear order in $D$. At $T=0$ for charged impurity scattering the long-wavelength limit of the dielectric function is \cite{Hwang_Gfn_Screening_PRB07}
\begin{equation}
\epsilon(q) = 1 + \frac{e^2}{4\pi \epsilon_0 \epsilon_r A} \, \bigg(\frac{k_F}{q}\bigg),
\end{equation}
yielding the Thomas-Fermi wave vector as $k_{TF} = e^2  k_F/(4\pi \epsilon_0 \epsilon_r A)$.


As a result, in topological insulators the matrix element $\bar{U}_{{\bm k}{\bm k}'}$ of a screened Coulomb potential between plane waves is given by
\begin{equation}\label{eq:W}
\arraycolsep 0.3ex
\begin{array}{rl}
\displaystyle \bar{U}_{{\bm k}{\bm k}'} = & \displaystyle \frac{Ze^2}{2\epsilon_0\epsilon_r}\, \frac{1}{ |{\bm k} - {\bm k}'| + k_{TF}} \\ [3ex] 
\displaystyle |\bar{U}_{{\bm k}{\bm k}'}|^2 = & \displaystyle \frac{Z^2e^4}{4\epsilon_0^2\epsilon_r^2}\, \bigg(\frac{1}{ |{\bm k} - {\bm k}'| + k_{TF}}\bigg)^2 \\ [3ex]
\equiv & \displaystyle \frac{W}{\big(\sin\frac{\gamma}{2} + \frac{k_{TF}}{2k_F}\big)^2},
\end{array}
\end{equation}
where $Z$ is the ionic charge (which we will assume to be $Z = 1$ below in this work) and $k_{TF}$ is the Thomas-Fermi wave vector. The parallel projection of the scattering term becomes
\begin{widetext}
\begin{equation}\label{zeta}
\arraycolsep 0.3ex
\begin{array}{rl}
\displaystyle P_\parallel \hat J(S_\parallel) = & \displaystyle \frac{kn_iW}{16\hbar \pi} \, \bigg[ \frac{1}{(A + 2Dk)} + \frac{1}{(A - 2Dk)} \bigg] \int d\theta' \, \frac{1 + \cos \gamma}{\big(\sin\frac{\gamma}{2} + \frac{k_{TF}}{2k_F}\big)^2} \, (s_\parallel - s'_\parallel) \, \sigma_{{\bm k}\parallel} \\ [3ex]
= & \displaystyle \frac{kn_i}{8\hbar \pi} \, \bigg[ \frac{1}{(A + 2Dk)} + \frac{1}{(A - 2Dk)} \bigg] \, \int d\theta' \, \zeta(\gamma) \, (s_\parallel - s'_\parallel)\, \sigma_{{\bm k}\parallel}  \\ [3ex] \displaystyle \zeta(\gamma) = & \displaystyle \frac{\cos^2\frac{\gamma}{2}}{\big(\sin\frac{\gamma}{2} + \frac{k_{TF}}{2k_F}\big)^2}.
\end{array}
\end{equation}
We expand $\zeta(\gamma)$ and $s_{{\bm k}\parallel} (\theta)$ in Fourier harmonics
\begin{equation}
\arraycolsep 0.3ex
\begin{array}{rl}
\displaystyle \zeta(\gamma) = & \displaystyle \sum_m \zeta_m \, e^{im\gamma} \\ [3ex] 
\displaystyle s_{{\bm k}\parallel} (\theta) = & \displaystyle \sum_m s_{\parallel m} \, e^{im\theta} \\ [3ex] 
\displaystyle P_\parallel \hat J(S_\parallel) = & \displaystyle \frac{kn_i}{4\hbar} \, \bigg[ \frac{1}{(A + 2Dk)} + \frac{1}{(A - 2Dk)} \bigg]  \sum_m (\zeta_0 - \zeta_m) \, s_{\parallel m} \, e^{- i m \theta} \, \, \sigma_{{\bm k}\parallel}, 
\end{array}
\end{equation}
\end{widetext}
and to obtain the expansion of $s_{{\bm k}\parallel} (\theta')$ replace $\theta \rightarrow \theta'$. The absence of backscattering, characteristic of topological insulators, is contained in the function $\zeta(\gamma)$. Equation (\ref{zeta}) shows that this function contains a factor of $(1 + \cos\gamma)$ which suppresses scattering for $\gamma = \pi$.

\section{Solution of the kinetic equation}
\label{sec:sol}

The Hamiltonian $\hat{H}_E$ describing the interaction with the electric field is given by $\hat{H}_E = e{\bm E}\cdot\hat{\bm r}$, with $\hat{\bm r}$ the position operator. It then follows straightforwardly that the kinetic equation takes the form
\begin{equation}
\td{f_{\bm k}}{t} + \frac{i}{\hbar}\, [H_{\bm k}, f_{\bm k}] + \hat{J}(f_{\bm k}) = \mathcal{D}.
\end{equation}
The driving term in the kinetic equation is $\mathcal{D} = \frac{e{\bm E }} {\hbar} \cdot \pd{\rho_{0{\bm k}}}{{\bm k}}$, where $\rho_{0{\bm k}}$ represents the equilibrium density matrix, given by $\rho_{0{\bm k}} = (f_{0{\bm k}+} + f_{0{\bm k}-}) \, \openone + (1/2)\, (f_{0{\bm k}+} - f_{0{\bm k}-}) \, {\bm\sigma}\cdot\hat{\bm k}$. The first term represents the charge density, while the second represents the spin density. We decompose the driving term into a scalar part $\mathcal{D}_n$ and a spin-dependent part $\mathcal{D}_s$, as 
\begin{equation}
\arraycolsep 0.3ex
\begin{array}{rl}
\displaystyle \mathcal{D}_n = & \displaystyle \frac{e{\bm E} \cdot \hat{\bm k}} {\hbar} \, (\pd{f_{0{\bm k}+}}{k} + \pd{f_{0{\bm k}-}}{k})  \, \openone \\ [3ex] 
\displaystyle \mathcal{D}_s = & \displaystyle \frac{1}{2}\, {\bm\sigma}\cdot\hat{\bm k}\, \frac{e{\bm E}\cdot\hat{\bm k}}{\hbar} \, (\pd{f_{0{\bm k}+}}{k} - \pd{f_{0{\bm k}-}}{k}) \\ [3ex]
+ & \displaystyle \frac{1}{2}\, {\bm\sigma}\cdot\hat{\bm \theta}\, \frac{e{\bm E}\cdot \hat{\bm \theta}}{\hbar k} (f_{0{\bm k}+} - f_{0{\bm k}-}).
\end{array}
\end{equation}
The latter is further decomposed into a part parallel to the Hamiltonian $\mathcal{D}_\parallel$ and a part perpendicular to it $\mathcal{D}_\perp$
\begin{equation}
\arraycolsep 0.3ex
\begin{array}{rl}
\displaystyle \mathcal{D}_{\parallel} = & \displaystyle \frac{1}{2} \, \frac{e{\bm E}\cdot\hat{\bm k}}{\hbar} \, (\pd{f_{0{\bm k}+}}{k} - \pd{f_{0{\bm k}-}}{k}) \, \sigma_{{\bm k}\parallel} = \frac{1}{2} \, d_{{\bm k}\parallel} \, \sigma_{{\bm k}\parallel} \\ [3ex] 
\displaystyle \mathcal{D}_{\perp} = & \displaystyle \frac{1}{2} \, \frac{e{\bm E}\cdot \hat{\bm \theta}}{\hbar k} \, (f_{0{\bm k}+} - f_{0{\bm k}-}) \, \sigma_{{\bm k}\perp} = \frac{1}{2} \, d_{{\bm k}\perp} \, \sigma_{{\bm k}\perp} .
\end{array}
\end{equation}
Evidently, since we consider electron doping, $f_{0{\bm k}-} = 0$. Nevertheless we retain $f_{0{\bm k}-} $ for completeness, since the term $(f_{0{\bm k}+} - f_{0{\bm k}-})$ will be shown to give rise to a contribution to the conductivity singular at the origin. 

We solve the kinetic equation. The first step is to divide it into three equations: one for the scalar part, one for the part parallel to the Hamiltonian and one for the part orthogonal to the Hamiltonian. 
\begin{subequations}\label{eq:Spp}
\begin{eqnarray}
\td{n_{\bm k}}{t} + P_n \hat J (f_{{\bm k}}) & = & \mathcal{D}_n, \\ [0.5ex]
\td{S_{{\bm k} \|}}{t} + P_\| \hat J (f_{{\bm k}}) & = & \mathcal{D}_\parallel, \\ [0.5ex]
\td{S_{{\bm k}\perp}}{t} + \frac{i}{\hbar} \, [H_{{\bm k}}, S_{{\bm k}\perp}] + P_\perp \hat J (f_{{\bm k}}) & = & \mathcal{D}_\perp.
\end{eqnarray}
\end{subequations}
We search for the solution as an expansion in the small parameter $\hbar/(\varepsilon_F\tau)$, where $\tau \propto n_i^{-1}$, to be defined below, represents the transport relaxation time. We label the orders in the expansion by superscripts with e.g. $S_{\bm k}^{(m)}$ representing $S_{\bm k}$ evaluated to order $m$. From these equations it is evident that the expansions of $n_{\bm k}$ and $s_{{\bm k}\parallel}$ begin at order $\varepsilon_F\tau/\hbar$, in other words $[\hbar/(\varepsilon_F\tau)]^{(-1)}$, whereas $s_{{\bm k}\perp}$ begins at order $[\hbar/(\varepsilon_F\tau)]^{(0)}$, in other words the leading term in it is independent of $\varepsilon_F\tau/\hbar$. Therefore, in determining the leading terms in $n_{\bm k}$ and $s_{{\bm k}\parallel}$, the coupling to $s_{{\bm k}\perp}$ can be neglected. We are then left with two coupled equations for $n_{\bm k}$ and $s_{{\bm k}\parallel}$, which are written as
\begin{equation}
\arraycolsep 0.3ex
\begin{array}{rl}
\displaystyle \td{n_{\bm k}}{t} + \hat J_{n \rightarrow n} (n_{\bm k}) + \hat J_{S \rightarrow n} (S_{\bm k}) = & \displaystyle \mathcal{D}_n \\ [3ex] 
\displaystyle \frac{1}{2}\, \td{s_{{\bm k}\parallel}}{t} \, \sigma_{{\bm k}\parallel} + P_\parallel \hat J_{S \rightarrow S}(S_\parallel) + P_\parallel \hat J_{n \rightarrow S} (n_{\bm k}) = & \displaystyle \frac{1}{2} \, d_{{\bm k}\parallel} \, \sigma_{{\bm k}\parallel}.
\end{array}
\end{equation}
We must carry out the projections in order to determine the scattering terms above. In the equations involving $P_\parallel$ we have eliminated $\sigma_{{\bm k}\parallel}$, since the equations for $S_{\bm k}$ have been cast in terms of equations for $s_{\bm k}$. To first order in $D$, the required scattering terms can be written as
\begin{equation}
\arraycolsep 0.3ex
\begin{array}{rl}
\displaystyle \hat J(n) \rightarrow & \displaystyle \frac{k n_i W}{4\pi\hbar A} \, \int d\theta' \, \zeta(\gamma) \big(n_{\bm k} - n_{{\bm k}'} \big) \\ [3ex]
\displaystyle \hat J_{S \rightarrow n} (S_{\bm k}) \rightarrow & \displaystyle - \frac{k n_i W Dk}{4\pi\hbar A^2} \, \int d\theta' \, \zeta(\gamma) \, (s_{{\bm k}\parallel} - s_{{\bm k}'\parallel}) \\ [3ex]
P_\parallel \displaystyle \hat J_{n \rightarrow S} (n_{\bm k}) \rightarrow & \displaystyle - \frac{k n_i W Dk}{2\pi\hbar A^2}\, \int d\theta'  \, \zeta(\gamma) \, \big(n_{\bm k} - n_{{\bm k}'} \big)\, \sigma_{{\bm k}\parallel}  \\ [3ex]
\displaystyle P_\parallel \hat J(S_{{\bm k}\parallel}) \rightarrow & \displaystyle \frac{k n_i W}{8\pi\hbar A} \, \int d\theta' \, \zeta(\gamma) \, (s_{{\bm k}\parallel} - s_{{\bm k}'\parallel}) \, \sigma_{{\bm k}\parallel}.
\end{array}
\end{equation}

The scattering terms have exactly the same structure. We proceed to solve the equations for $n_{\bm k}$ and $s_{{\bm k}\parallel}$. We note that the RHS on both equations in this case has only Fourier component 1, so we can make the Ansatz $n_{\bm k} = n_1 \, e^{- i \theta} + n_{-1} \, e^{i \theta}$  and $s_{{\bm k}\parallel} = s_{\parallel 1} \, e^{- i \theta} + s_{\parallel -1}\, e^{i \theta}$. Observing in addition that $\mathcal{D}_n = d_{{\bm k}\parallel}$ allows us to write the coupled equations in the steady state as
\begin{equation}
\arraycolsep 0.3ex
\begin{array}{rl}
\displaystyle \frac{2n_{\bm k}}{\tau} - \frac{s_{{\bm k}\parallel}}{\tau_D} = & \displaystyle \mathcal{D}_n \\ [3ex]
\displaystyle - \frac{2n_{\bm k}}{\tau_D} + \frac{s_{{\bm k}\parallel}}{\tau} = & \displaystyle \frac{\mathcal{D}_n}{2},
\end{array}
\end{equation}
where $\tau$, the transport relaxation time, is given by the expression
\begin{equation}\label{tau}
\displaystyle \frac{1}{\tau} = \frac{kn_iW}{2\hbar A} \, (\zeta_0 - \zeta_1),
\end{equation}
and we have also introduced
\begin{equation}\label{tauD}
\displaystyle \frac{1}{\tau_D} = \frac{kn_iWDk}{\hbar A^2} \, (\zeta_0 - \zeta_1),
\end{equation}
as a measure of the coupling of the scalar and spin distributions by scattering. The equations are trivial to solve and we obtain in the steady state
\begin{equation}
\arraycolsep 0.3ex
\begin{array}{rl}
\displaystyle n_{\bm k} = & \displaystyle \frac{\mathcal{D}_n \tau}{2} \, (1 + \tau/2\tau_D) \\ [3ex]
\displaystyle s_{{\bm k}\parallel} = & \displaystyle \frac{\mathcal{D}_n \tau}{2} \, (1 + 2\tau/\tau_D).
\end{array}
\end{equation}
The equation is valid to first-order in $\tau/\tau_D$. Note that, as $D \rightarrow 0$, $\tau_D \rightarrow \infty$ and we recover the results for the bare Dirac cone.

Next we examine the term $S_{{\bm k}\perp}^{(0)}$. The kinetic equation for $S_{{\bm k}\perp}^{(0)}$ is
\begin{equation}
\arraycolsep 0.3ex
\begin{array}{rl}
\displaystyle \td{S_{{\bm k}\perp}^{(0)}}{t} + \frac{i}{\hbar} \, [H_{{\bm k}}, S_{{\bm k}\perp}^{(0)}] = & \displaystyle d_\perp - P_\perp \hat J[n_{{\bm k} \|}^{(-1)}]  - P_\perp \hat J[S_{{\bm k} \|}^{(-1)}].
\end{array}
\end{equation}
The projections of the scattering terms do not contribute to the conductivity. The bare source term $d_\perp$ is 
\begin{equation}
d_\perp = \frac{e{\bm E}\cdot \hat{\bm \theta}}{\hbar k} \, (f_{0{\bm k}+} - f_{0{\bm k}-}).
\end{equation}
The result for $S_{{\bm k}\perp}^{(0)}$ is
\begin{widetext}
\begin{equation}
S_{{\bm k}\perp}^{(0)} = \frac{e{\bm E}\cdot \hat{\bm \theta}}{2\hbar k} \, (f_{0{\bm k}+} - f_{0{\bm k}-}) \int_0^\infty dt \, e^{-\eta t} \Bigl(\sigma_{{\bm k} \perp} \, \cos \frac{2Akt}{\hbar} + {\bm \sigma}\cdot\hat{\bm k}\times\hat{\bm \theta} \, \sin \frac{2Akt}{\hbar}\Bigr).
\end{equation}
\end{widetext}
The part $\propto {\bm \sigma}\cdot\hat{\bm k}\times\hat{\bm \theta}$ averages to zero over directions in momentum space and is not considered any further.
Integrating the cosine over time gives
\begin{equation}\label{eq:Sperp3}
S_{{\bm k}\perp}^{(0)} = \frac{e{\bm E}\cdot \hat{\bm \theta}}{2\hbar k} \, (f_{0{\bm k}+} - f_{0{\bm k}-}) \, \frac{\eta}{\frac{4A^2k^2}{\hbar^2} + \eta^2} \, \sigma_{{\bm k}\perp}
\end{equation}
 
\subsection{Current operator and conductivity}

The current density operator is decomposed into a parallel and a perpendicular part, the electric field is assumed $\parallel\hat{\bm x}$ and we are looking
for the diagonal conductivity
\begin{equation}
\arraycolsep 0.3ex
\begin{array}{rl}
\displaystyle j_i ({\bm k}) = & \displaystyle j_{i,n} + \frac{1}{2}\, j_{i, \parallel} \, \sigma_\parallel + \frac{1}{2}\, j_{i, \perp} \, \sigma_\perp.
\end{array}
\end{equation}
The three contributions to the current operator, representing the scalar part, the part parallel to the Hamiltonian and the part orthogonal to the Hamiltonian, are given explicitly by
\begin{equation}
\arraycolsep 0.3ex
\begin{array}{rl} 
\displaystyle j_{x, n} = & \displaystyle - \frac{2e D k}{\hbar} \, \cos\theta \\ [3ex] 
\displaystyle j_{x, \parallel} = & \displaystyle - \frac{2e A}{\hbar} \, \cos\theta \\ [3ex] 
\displaystyle j_{x, \perp} = & \displaystyle \frac{2e A}{\hbar} \, \sin\theta .
\end{array}
\end{equation}
Using this form for the current operator it is trivial to show that the conductivity linear in $\tau$, keeping only terms linear in $D$, is given by
\begin{equation}\label{eq:ord}
\sigma^\mathrm{Boltz}_{xx} = \frac{e^2}{h} \, \bigg[\frac{A \, k_F \, \tau}{4\hbar}\, \bigg(1 + \frac{2\tau}{\tau_D} \bigg) + \frac{D \, k_F^2 \, \tau}{4\hbar} \bigg],
\end{equation}
We emphasize again that this result tends to the result for graphene\cite{SDS_Gfn_RMP10, Culcer_Gfn_Transp_PRB08} when $D \rightarrow 0$ (note that the definition of $\tau$ in Refs.\ \onlinecite{SDS_Gfn_RMP10, Culcer_Gfn_Transp_PRB08} differs by a factor of 2.) The contribution to the conductivity independent of $\tau$ is
\begin{equation}\label{eq:min}
\arraycolsep 0.3ex
\begin{array}{rl}
\displaystyle \sigma^{0\perp}_{xx} = & \displaystyle \bigg(\frac{\pi e^2}{8h}\bigg)  \,  \lim_{\omega \rightarrow 0} \, \biggl[ \frac{1}{1 + e^{\beta(\mu + \hbar\omega/2)}} - \frac{1}{1 + e^{\beta(\mu - \hbar\omega/2)}} \biggr].
\end{array}
\end{equation}
The latter term however is zero for finite $\mu$, which is the case we are considering. Thus, the leading order term in the conductivity is $\propto \tau$, and in the limit $\varepsilon_F\tau/\hbar \gg 1$ there is no term of order $\tau^0$.

\section{Implications for transport experiments}
\label{sec:expt}

In this section, as in the calculation of screening above, it is assumed the system has a low enough density that the Dirac-cone approximation is applicable, which for Bi$_2$Se$_3$ implies $k_F < 10^{8}$m$^{-1}$. Below we use the terminology \textit{high density} as meaning that the density is high enough that the $Dk^2$ term in the Hamiltonian has a noticeable effect (meaning $n \approx 10^{13}$cm$^{-2}$ and higher), with \textit{low density} reserved for situations in which this term is negligible. Nevertheless we retain the scattering time $\tau_D$, which allows us to examine the correction to the low-density expressions when these terms become noticeable.

\subsection{Density dependence}

The conductivity is a function of two main parameters accessible experimentally: the carrier number density and the impurity density/scattering time. The dependence of the conductivity on the carrier number density arises through its direct dependence on $k_F$ and through its dependence on $n$ through $\tau$. In terms of the number density the Fermi wave vector is given by $k_F^2 = 4\pi n$. The transport time is given by Eq.\ (\ref{tau}) and $W$ is defined in Eq.\ (\ref{eq:W}). The number-density dependence of $\tau$ depends on the dominant form of scattering and whether the number density is high or low as defined above. For charged impurity scattering $W \propto k^{-2}$, so that $\tau \propto k$ and the two terms in the conductivity are $\propto n$ and $n^{3/2}$. At the same time, in a two-dimensional system surface roughness gives rise to short-range scattering as discussed in Ref. \onlinecite{Ando_RMP82}. For short range scattering of this type $W$ is a constant and $\tau \propto k$, with the two terms in the conductivity being a constant and $n^{1/2}$. These results are summarized in Table \ref{tab:scattering}. At high density the term $\propto Dk$ becomes noticeable through the scattering time $\tau_D$. Since $\tau_D$ has an extra factor of $k_F$, retaining terms to first order in $D$ gives us an additional term $\propto Dn^{1/2}$, resulting in a further contribution to the conductivity of the form $\propto D n^{3/2}$, as is manifest in Eq.\ (\ref{eq:ord}). Taking this fact into account, in order to keep the expressions simple, in Table \ref{tab:scattering} we have listed only the number density dependence explicitly, replacing the constants of proportionality by generic constants.

\begin{table}[tbp] 
  \caption{\label{tab:scattering} Carrier density dependence of the conductivity. High-density and low-density regimes for scattering by screened charged impurities and for short-range impurities. The numbers $a_1 - a_4$ represent constants.}
  $\arraycolsep 1em
   \begin{array}{c@{\hspace{2em}}cc} \hline\hline
     & {\rm Screened \,\, charges} & {\rm Short-range} \\ \hline
         {\rm Low \,\, density}   & a_1n & a_3 \\ 
         {\rm High \,\, density}  & a_1n + a_2n^{3/2} & a_3 + a_4 n^{1/2} \\ \hline\hline
  \end{array}$
\end{table}

\subsection{Scattering times}

Two scattering times commonly encountered in momentum relaxation: the transport relaxation time $\tau$ and the quantum lifetime $\tau_Q$. The former represents momentum relaxation in transport and is weighted by a factor of $(1 - \cos\gamma)$ to reflect the fact that forward scattering does not alter the charge current. This is the time measured in transport experiments. The latter represents the time taken to scatter from one Bloch state into another, thus all scattering amplitudes are weighted equally. This time determines observables such as the Dingle temperature and the broadening of Shubnikov-de Haas oscillations. As a result of the different weighting of the scattering amplitudes in $\tau$ and $\tau_Q$, the ratio $\tau/\tau_Q$ characterizes the scattering by providing information on which angles are dominant.\cite{Hwang_Gfn_tstp_PRB08} We determine the ratio of these two scattering times in topological insulators. First, we provide expressions for $\tau$ and $\tau_Q$ in the presence of charged impurity scattering. To begin with, we write the quantum lifetime $\tau_Q$ and the transport lifetime $\tau$ in a slightly different way from the way they were written in Sec. \ref{sec:sol} (with $D=0$, and $d=0$):
\begin{equation}
\arraycolsep 0.3ex
\begin{array}{rl} 
\displaystyle \frac{1}{\tau} = & \displaystyle \frac{k_Fn_iW}{4\pi\hbar A} \, \int_0^{2\pi} d\gamma \, \zeta(\gamma) \, (1 - \cos\gamma) \\ [3ex]
\displaystyle \frac{1}{\tau_Q} = & \displaystyle \frac{k_Fn_iW}{4\pi\hbar A} \, \int_0^{2\pi} d\gamma \, \zeta(\gamma),
\end{array}
\end{equation}
where $\tau_Q$ is discussed at length in Ref.\ \onlinecite{Hwang_Gfn_tstp_PRB08}.
\begin{table}[tbp]
  \caption{\label{tab:ratio} Ratio of the transport and Bloch scattering times $\tau/\tau_Q$ for $r_s \ll 4$, $r_s \gg 4$ and different scattering mechanisms. $b_1 - b_8$ are constants.}
  $\arraycolsep 1em
   \begin{array}{c@{\hspace{2em}}cc} \hline\hline
     & {\rm Screened \,\, charges} & {\rm Short-range} \\ \hline
         r_s \ll 4 & b_1 r_s^{-1} - b_2 \ln r_s & b_5 + b_6 r_s \ln r_s \\ 
         r_s \gg 4 & b_3 + b_4 r_s^{-1} & b_7 + b_8 r_s \\ \hline\hline
  \end{array}$
\end{table}
We can also express them in terms of the original scattering potential $\bar{U}_{{\bm k}{\bm k}'} \equiv \bar{U}(q)$ as
\begin{equation}
\arraycolsep 0.3ex
\begin{array}{rl} 
\displaystyle \frac{1}{\tau} = & \displaystyle \, \frac{k_Fn_i}{4\pi\hbar A} \, \int_0^\pi d\gamma \, \bigg[\frac{\bar{U}(q)}{\epsilon(q)}\bigg]^2 \, (1 - \cos^2\gamma) \\ [3ex] 
\displaystyle \frac{1}{\tau_Q} = & \displaystyle \, \frac{k_Fn_i}{4\pi\hbar A} \, \int_0^\pi d\gamma \, \bigg[\frac{\bar{U}(q)}{\epsilon(q)}\bigg]^2 \, (1 + \cos\gamma).
\end{array}
\end{equation}
We recall that for elastic scattering of carriers on the Fermi surface $|{\bm k} - {\bm k}'| \equiv q \rightarrow 2k_F \sin\gamma/2$. We consider first charged impurity scattering, denoted by the subscript $c$. The two scattering times can be expressed in terms of a prefactor with the units of frequency multiplying a dimensionless angular integral, thus $1/\tau_{Qc} = (r_s^2/\tau_{0c})\, I_{Qc}$ and $1/\tau_c = (r_s^2/\tau_{0c})\, I_{tc}$, where the Wigner-Seitz radius $r_s = e^2/(2\pi \epsilon_0\epsilon_r A)$, the quantity 
\begin{equation}
\frac{1}{\tau_{0c}} = \frac{A n_i}{32\hbar}\sqrt{\frac{\pi}{n}} 
\end{equation}
and the angular integrals $I_t$ and $I_Q$
\begin{equation}
\arraycolsep 0.3ex
\begin{array}{rl}
\displaystyle I_{tc} (x) = & \displaystyle \int_0^\pi d\gamma \, \frac{1 - \cos^2\gamma}{(\sin\frac{\gamma}{2} + x)^2} \\ [3ex]
\displaystyle I_{Qc} (x) = & \displaystyle \int_0^\pi d\gamma \, \frac{1 + \cos\gamma}{(\sin\frac{\gamma}{2} + x)^2}.
\end{array}
\end{equation}
Even before evaluating the integrals explicitly it will be observed that the transport time $\tau$ is not sensitive to either small or large angle scattering, while the quantum lifetime $\tau_Q$ is insensitive to large angle scattering. The qualitatively different dependence on the scattering angles makes plain the fact that the ratio $\tau/\tau_Q$ will have a qualitatively different behavior as a function of $r_s$ for charged impurity scattering and for short-range scattering. It must be emphasized also that, despite the superficial similarity to graphene, the behavior of this ratio as a function of $r_s$ for real materials is also quite different, due to the absence of the spin and valley degeneracy, and the fact that $r_s$ for a topological insulator differs from its value in graphene (for the same parameters) by an overall factor of 2.

Substituting for the screening function we can determine the expressions for these times as a function of $r_s$ in the limits of small and large $r_s$. The results are qualitatively the same as in Ref.\ \onlinecite{Hwang_Gfn_tstp_PRB08} and are summarized in Table \ref{tab:ratio}. For charged impurity scattering, we find for the integral $I_{Qc}$
\begin{widetext}
 \begin{equation}
\arraycolsep 0.3ex
\begin{array}{rl} 
\displaystyle I_{Qc} (x) = & \displaystyle \frac{4}{x} - 2\pi + 8x \, \frac{(2 - x^2) {\rm arccoth} \sqrt{1 - x^2} - (1 - x^2) {\rm arccoth} \sqrt{\frac{1 - x}{1 + x}})}{(1 - x^2)^{3/2}} , x < 1 \\ [3ex]
 & \displaystyle 8 - 2 \pi, x =1 \\ [3ex]
 & \displaystyle -2\pi + \frac{4}{x} - \frac{8x}{\sqrt{x^2 - 1}} \, \bigg( {\rm arccot} \sqrt{x^2 - 1} - \arctan \sqrt{\frac{x + 1}{x - 1}}  \bigg) , x >1.
\end{array}
\end{equation}
while $I_{tc}$ can be written as
\begin{equation}
\arraycolsep 0.3ex
\begin{array}{rl} 
\displaystyle I_{tc} (x) = & \displaystyle 2 (\pi + 12x - 6\pi x^2) + \frac{16x(3x^2 - 2)}{\sqrt{1 - x^2}} \, \bigg( {\rm arccoth} \sqrt{1 - x^2} - {\rm arctanh}\sqrt{\frac{1 + x}{1 - x}}\bigg), x < 1 \\ [3ex]
 & \displaystyle 32 - 10 \pi, x =1 \\ [3ex]
 & \displaystyle 2 (\pi + 12x - 6\pi x^2) - \frac{16x(3x^2 - 2)}{\sqrt{x^2 - 1}} \bigg({\rm arccot} \sqrt{x^2 - 1} -  \arctan \sqrt{\frac{x + 1}{x - 1}} \bigg), x >1
 \end{array}
\end{equation} 
\end{widetext}
Setting $x = r_s/4$ we find that the ratio of the transport and quantum lifetimes for charged impurity scattering in the limit $r_s \ll 4$ can be expressed as
\begin{equation}\label{ratioc}
\bigg(\frac{\tau}{\tau_Q}\bigg)_{c} = \frac{8}{\pi r_s} - \frac{16 \ln r_s}{\pi^2} + ... 
\end{equation}
For short-range scatterers characterized by a potential $v_0\delta({\bm r})$, the transport and quantum scattering times are given by the following expressions
\begin{equation}\label{ShortRangeTimes}
\arraycolsep 0.3ex
\begin{array}{rl}
\displaystyle \frac{1}{\tau_\delta} = & \displaystyle \frac{1}{\tau_{0\delta}} I_{t\delta} \big( \frac{r_s}{4} \big) \\ [3ex]
\displaystyle \frac{1}{\tau_{Q\delta}} = & \displaystyle \frac{1}{\tau_{0\delta}} I_{Q\delta} \big( \frac{r_s}{4} \big),
\end{array} 
\end{equation}
where the quantity $\tau_{0\delta}$ is defined by
\begin{equation}
\frac{1}{\tau_{0\delta}} = \frac{n_iv_0^2\sqrt{n}}{2A\sqrt{\pi} \hbar}
\end{equation}
and the integrals appearing in Eqs.\ (\ref{ShortRangeTimes}) are
\begin{equation}
\begin{array}{rl}
\displaystyle I_{t\delta} (x) = & \displaystyle \int_0^\pi d\gamma \bigg(\frac{\sin\frac{\gamma}{2}}{\sin\frac{\gamma}{2} + x} \bigg)^2 \, (1 - \cos^2\gamma) \\ [3ex]
\displaystyle I_{Q\delta} (x) = & \displaystyle \int_0^\pi d\gamma \bigg(\frac{\sin\frac{\gamma}{2}}{\sin\frac{\gamma}{2} + x} \bigg)^2 \, (1 + \cos\gamma). 
\end{array} 
\end{equation}
Evaluating the integrals explicitly yields
\begin{widetext}
\begin{equation}
\begin{array}{rl}
\displaystyle I_{Q\delta} (x) = & \displaystyle \pi + 12\pi x - 6\pi x^2 + \frac{8x(3x^2 - 2)({\rm arccoth}\sqrt{1 - x^2} - {\rm arccoth}\sqrt{\frac{1 - x}{1 + x}})}{\sqrt{1 - x^2}}, x < 1 \\ [3ex]
& \displaystyle 16 - 5\pi, x = 1 \\ [3ex]
& \displaystyle \pi + 12\pi x - 6\pi x^2 - \frac{8x(3x^2 - 2)({\rm arccoth}\sqrt{x^2 - 1} - {\rm arccoth}\sqrt{\frac{x - 1}{x + 1}})}{\sqrt{1 - x^2}}, x > 1,
\end{array} 
\end{equation}
and
\begin{equation}
\begin{array}{rl}
\displaystyle I_{t\delta} (x) = & \displaystyle \frac{\pi}{2} - \frac{16 x}{3} + 6 \pi x^2 + 40 x^3 - 20 \pi x^4 + \frac{16 x^3 (5x^2 - 4)({\rm arccoth}\sqrt{1 - x^2} - {\rm arccoth}\sqrt{\frac{1 - x}{1 + x}})}{\sqrt{1 - x^2}}, x < 1 \\ [3ex]
& \displaystyle \frac{1}{6} \, (256 - 81 \pi), x = 1 \\ [3ex]
& \displaystyle \frac{\pi}{2} - \frac{16 x}{3} + 6 \pi x^2 + 40 x^3 - 20 \pi x^4 - \frac{16 x^3 (5x^2 - 4)({\rm arccot}\sqrt{x^2 - 1} - {\rm arccot}\sqrt{\frac{x - 1}{x + 1}})}{\sqrt{x^2 - 1}}, x > 1. 
\end{array} 
\end{equation}
\end{widetext}
Setting again $x = r_s/4$ the ratio of the transport and quantum lifetimes in the limit $r_s \ll 4$ for short-range impurities takes the form
\begin{equation}\label{ratiod}
\bigg(\frac{\tau}{\tau_Q}\bigg)_{\delta} = 2 + \frac{4r_s \ln r_s}{\pi} + \frac{2(17 - 18 \ln 2)r_s}{3\pi} + ... 
\end{equation}
In the range covered here this ratio does not exceed 2 for short-range scatterers. We would also like to point out that $r_s \propto 1/A$ for the surface of Bi$_2$Te$_3$ is approximately twice that in graphene, i.e. $r_s \approx 0.14$, which is $\ll 4$. \footnote{The results of Eq. (\ref{ratioc}) and Eq. (\ref{ratiod}) can be obtained by substituting $r_s/8$ into Eqs. (16) and (19) of Ref.\ \onlinecite{Hwang_Gfn_tstp_PRB08}.}

\subsection{Numerical results}

\begin{figure}[tbp]
\includegraphics[width=\columnwidth]{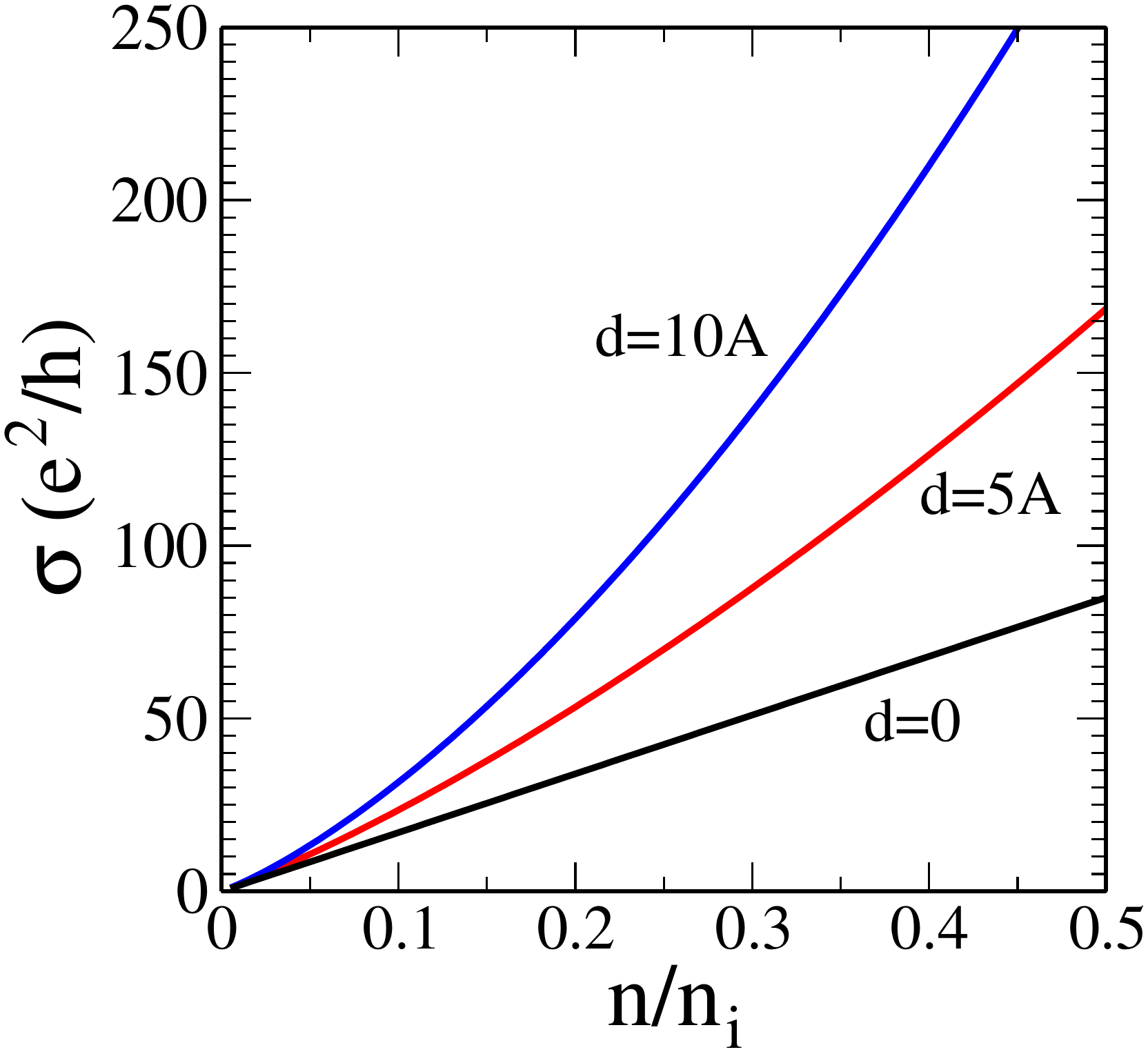}
\caption{\label{sigma}
Calculated conductivity limited by screened charged impurities as
  a function of the surface carrier density for different impurity
  locations $d=0$, 5, 10\AA. In this figure we use the
  following parameters: an impurity density $n_i=10^{13}$ cm$^{-2}$,
  $A=4.1$ eV\AA, which corresponds to the Fermi velocity
  $v_F=6.2\times 10^7$ cm/s, and the static dielectric constant $\epsilon_r=100$. 
}
\end{figure}

We conclude our discussion with numerical results on the conductivity and the ratio of the two scattering times. From the transport scattering time calculated in the previous section we can calculate the conductivity using Eq. (35) for $d=0$. In Fig.\ \ref{sigma} we have plotted the conductivity versus the ratio $n/n_i$ of the carrier number density to the impurity number density for several values of $d$, the distance of the impurities from the surface of the topological insulator. For $d=0$, when the impurities are located right on the surface, the conductivity is a linear function of $n/n_i$. If $d$ is large, meaning $k_F d > 1$ then large angle scattering is strongly suppressed. This leads to a large increase in $\tau/\tau_Q$, since the Fourier transform of the scattering potential is suppressed by $e^{-qd}$. 

\begin{figure}[tbp]
\bigskip
\includegraphics[width=\columnwidth]{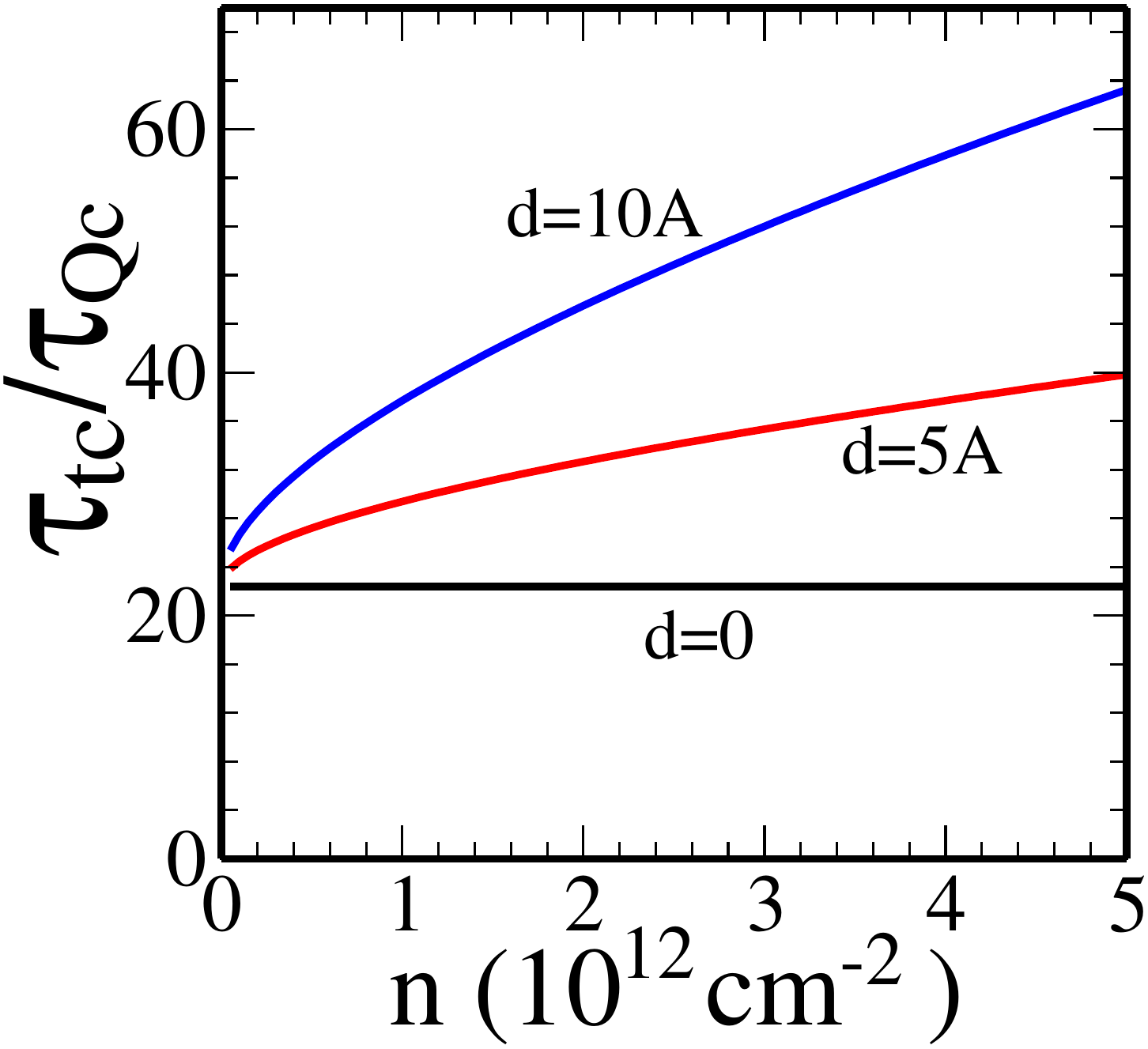}
\caption{\label{ratio}
The ratio of the transport relaxation time to the quantum life time
for screened charged impurities as
a function of the carrier density for different impurity locations.
We use the same parameters as Fig. \ref{sigma}.
The ratio for the short-range impurity is independent of
carrier density and we find $\tau_{t\delta}/\tau_{Q\delta}=1.46$.
}
\end{figure}

In Fig.\ \ref{ratio} we have plotted the ratio of the two scattering times as a function of the carrier number density for charged impurity scattering for different values of the distance $d$ of the impurities from the surface. For $d=0$ this ratio is not a function of $n$. It should also be noted that, for short range scatterers, the ratio $\tau_\delta/\tau_{Q\delta}$ is a constant as a function of $n$, which we find to be 1.46.

We note that the analytical results determined in the previous sections for $d=0$ agree with the numerical results shown in Figs. 1 and 2. Nevertheless, for $d \neq 0$, it is not possible to compare directly the numerical results with the analytical results, the latter being valid for $d=0$, i.e. the impurities located on the surface.

Ref. \onlinecite{Butch_Bi2Se3_Search_10} measured a mobility of the order of 1 m$^2$/(Vs) at a number density of the order of $\approx 10^{17}cm^{-3}$. For example, for one sample the surface carrier density is estimated at $\approx 7 \times 10^{12}$cm$^{-2}$, while the effective scattering rate is 0.8ps, which is obtained from transport. For such a sample $\varepsilon_F\tau/\hbar \approx 35$, placing it within the range of applicability of our theory. However the experiment is not sensing surface state transport so a direct comparison is not feasible at present.

Our theory has focused on scattering due to charged impurities and interface roughness, whereas in a realistic sample other scattering mechanisms may exist, primarily phonons. Phonons can be included easily in our theory as an additional contribution to the scattering term. However in this work we have considered zero temperature, therefore there is no phonon absorption and there is only phonon emission, which usually has a negligible contribution to the conductivity.

We note that our theoretical results would be useful in discerning transport by the surface topological 2D states only after a real topological insulator material, i.e. a system which is a bulk insulator, becomes available. All currently existing TI materials are effectively bulk \textit{metals} because of their large unintentional doping. Such systems are, by definition, unsuitable for studying surface transport properties since bulk transport completely overwhelms any surface transport signature. Discussing TI surface transport in such bulk-doped TI materials is not particularly meaningful since it must necessarily involve complex data fitting and assumptions in order to distinguish bulk versus surface transport contributions. Although such analyses are being carried out in bulk doped TI materials, in order to understand the transport behavior (and our theoretical results should certainly help such analyses), we believe that any real progress will come only when surface TI transport can be carried out unambiguously, without any complications arising from the (more dominant) bulk transport channel. We believe that our theoretical results presented in this paper would become important in the context of such surface TI studies in the laboratory.

What we have shown in this paper is that the so-called topological protection of TI surface transport is a protection only against localization by back scattering, not a protection against impurity or defect scattering. The presence of impurities and defects will certainly lead to scattering of the TI surface carriers and the surface 2D conductivity will be strongly affected by such scattering. If such scattering is strong (which will be a non-universal feature of the sample quality), then the actual surface resistivity will be very high and the associated mobility very low. There is no guarantee or protection against low carrier mobility in the TI surface states whatsoever, and unless one has very clean surfaces, there is little hope of studying surface transport in 2D topological states, notwithstanding their observation in beautiful band structure measurements through ARPES or STM experiments.

\subsection{Minimum conductivity at the Dirac point}

Thus far we have concentrated on the conductivity $\sigma_{xx}$ at carrier densities such that $\varepsilon_F \tau/\hbar \gg 1$. This is the limit that is immediately relevant to experiment given the high unintentional doping in topological insulators. At such densities the conductivity is given by $\sigma^{Boltz}_{xx}$ from Eq.\ (\ref{eq:ord}), while the term $\sigma^{0\perp}_{xx}$ from Eq.\ (\ref{eq:min}) does not contribute. At present it is not possible to tune the number density continuously through the Dirac point as is done in graphene. Nevertheless, given the similarity of the Hamiltonians of graphene and topological insulators, we expect a similar behavior at the Dirac point. We predict that, when such tuning does become possible, a minimum conductivity plateau around the Dirac point will be observed, analogously to graphene.  

The term $\sigma^{0\perp}_{xx}$ is expected to provide one contribution to the minimum conductivity, which can also be obtained using a field-theoretical method in the absence of disorder. \cite{SDS_Gfn_RMP10} This method can in fact yield several values, depending on the choice of regularization procedure, which are close in magnitude. However, as $n \rightarrow 0$ the condition $\varepsilon_F \tau/\hbar \gg 1$ is inevitably violated, and a strong renormalization of $\sigma^{0\perp}_{xx}$ is expected due to charged impurities.\cite{Adam_Gfn_PNAS07, Rossi_Gfn_RandCrgImp_PRL08, Rossi_Gfn_EffMdm_PRB09} Quite generally, charged impurities give rise to an inhomogeneous Coulomb potential, which is screened by \textit{both} electrons and holes. The net effects of this potential are an inhomogeneity in the carrier density itself and a shift in the Dirac point as a function of position. At high densities the spatial fluctuations in the carrier density are of secondary importance, and do not modify the linear dependence of the conductivity on $n$, yet as the chemical potential approaches the Dirac point, where the average carrier density $\bkt{n} = 0$, these fluctuations play the dominant role in conduction.\cite{Adam_Gfn_PNAS07} At low densities, the carriers cluster into puddles of electrons \textit{and} holes, and a residual density of carriers is always present, making it impossible to reach the Dirac point experimentally. \cite{Adam_Gfn_PNAS07} Consequently, although Eq.\ (\ref{eq:min}) appears to give a universal value for the minimum conductivity, the renormalization due to the presence of electron and hole puddles displays a strong sample dependence: in graphene it results in an enhancement of $\sigma^{0\perp}_{xx}$ by a factor of 2-20.\cite{SDS_Gfn_RMP10, Rossi_Gfn_RandCrgImp_PRL08} In addition, exchange and correlation effects make a significant contribution to the minimum conductivity. \cite{Rossi_Gfn_EffMdm_PRB09}

An accurate determination of this enhancement for topological insulators requires detailed knowledge of the impurity density distribution, which can only be determined experimentally or modeled numerically by means of e.g. an effective medium theory. \cite{Rossi_Gfn_EffMdm_PRB09} However, a self consistent transport theory provides a physically transparent way to identify the approximate minimum conductivity. The magnitude of voltage fluctuations can be calculated in the random phase approximation, and the result used to determine the residual density and the critical number density at which the transition occurs to the regime of electron and hole puddles, where the carrier density will be highly inhomogeneous. Both the residual density and the critical density are proportional to $n_i$ by a factor of order unity, \cite{Adam_Gfn_PNAS07} and as a first approximation one may take both of them as $\approx n_i$. In view of these observations, the minimum conductivity plateau will be seen approximately at $\sigma_{xx}^{Boltz}$, in which with the carrier density $n$ is replaced by $n_i$. Using the expressions found above for $\sigma_{xx}^{Boltz}$ and $\tau$, this yields for topological insulators
\begin{equation}
\sigma_{xx}^{min} \approx \frac{e^2}{h} \, \bigg(\frac{8}{I_{tc}}\bigg).
\end{equation}
At carrier densities approaching zero the $D$-dependent term in the Hamiltonian will not contribute. Therefore, when experiment will be able to tune the carrier density to zero we expect a minimum conductivity of the order of $\sigma^{min}_{xx}$ to be observed. We note that, with $r_s$ much smaller than in graphene due to the large dielectric constant, the minimum conductivity may be substantially larger. \cite{Rossi_Gfn_EffMdm_PRB09}

\subsection{Current-induced spin polarization}

The charge current is proportional to the spin operator, as can be seen from Eq.\ (\ref{j}). Therefore a nonzero steady-state surface charge current automatically translates into a nonzero steady-state surface spin density. The spin density can easily be found by simply multiplying the charge current by $\hbar^2/(-2eA)$, yielding
\begin{equation}
s_x = -\frac{eE_x}{4\pi} \, \bigg[\frac{A \, k_F \, \tau}{4\hbar}\, \bigg(1 + \frac{2\tau}{\tau_D} \bigg) + \frac{D \, k_F^2 \, \tau}{4\hbar} \bigg],
\end{equation}
Since ${\bm \sigma}$ in the Hamiltonian represents the rotated spin, this corresponds to a net density of $s_y$. The effective Hamiltonian of Eq. (\ref{Ham}) is obtained from the original Rashba Hamiltonian by replacing $\sigma_y \rightarrow \sigma_x$ and $\sigma_x \rightarrow - \sigma_y$. For a sample in which the impurities are located on the surface, with $n/n_i = 0.5$ and $E_x = 25000V/m$, the spin density is $\approx 5 \times 10^{14}$ $spins/m^2$ (where spin $\equiv \hbar$), which corresponds to approximately $10^{-4}$ spins per unit cell area. This number, although small, can be detected experimentally using a surface spin probe such as Kerr rotation. It is also a conservative estimate: for very clean samples, having either a smaller impurity density or impurities located further away from the surface, this number can reach much higher values. Current-induced spin polarization is a signature of two-dimensional transport in topological insulators since there is no spin polarization from the bulk.

\section{Effect of metallic contacts}
\label{sec:contacts}

A question of major practical importance concerns the stability of the surface states in the presence of metallic contacts. Transport measurements require the addition of metallic contacts on the surface of the topological insulator. Since the properties of the surface states depend crucially on the boundary conditions, the natural question is how are these properties affected by the metallic contacts. To address this question, we consider the case of large area contacts and perform numerical calculations for the minimal tight-binding model studied in Ref. \onlinecite{Stanescu_TI_Proximity_10}. A topological insulator in contact with a metal along a planar interface is described by the Hamiltonian 
\begin{equation}
H_{\rm tot}=H_{\rm TI}+H_{\rm M}+H_{\rm t}. \label{Htot}
\end{equation}
The topological insulator term $H_{\rm TI}$ is given by the tight binding model on a diamond lattice with spin-orbit interactions\cite{Fu_3DTI_PRL07}
\begin{align}
H_{\rm TI}\!=\! t\!\sum_{\langle ij\rangle, \sigma}  c_{i\sigma}^\dagger c_{j\sigma}
\!+\! i\lambda_{SO}\! \sum_{\langle\langle ij \rangle\rangle}\!
 {\bf S}_{\sigma,\sigma'}\! \cdot \! ({\bf d}_{ij}^1 \! \times \! {\bf d}_{ij}^2) c_{i\sigma}^\dagger c_{j\sigma'}, 
 \label{eq:tbmodel}
\end{align}
where the first term is the nearest neighbor hopping on the diamond lattice and  the second term connects second order neighbors with a spin and direction dependent amplitude. The direction dependence is given by the bond vectors ${\bf d}_{ij}^{1,2}$  traversed between sites $i$ and $j$. The metallic term in Eq. (\ref{Htot}) is given by
\begin{align}\label{eq:scmodel}
H_{\rm M}= \sum_{\langle ij\rangle, \sigma} t_{ij}^{\prime} a_{i\sigma}^\dagger a_{j\sigma}.
\end{align}
The model is defined on a simple hexagonal lattice with a lattice constant for the triangular basis that ensures simple interface matching conditions between the metal and the ($1,1,1$) surface of the TI.  The tunneling term in Eq.   (\ref{Htot}) is
\begin{align}
H_{\rm t}=\sum_{\langle ij \rangle }  \tilde t (a_{i\sigma}^\dagger c_{j\sigma}+c^\dag_{j\sigma} a_{i\sigma}), \label{Ht}
\end{align}
were $\tilde t$ is the tunneling matrix element that characterizes the transparency of the interface between the topological insulator and the  metal. 

\begin{figure}[tbp]
\begin{center}
\includegraphics[width=0.45\textwidth]{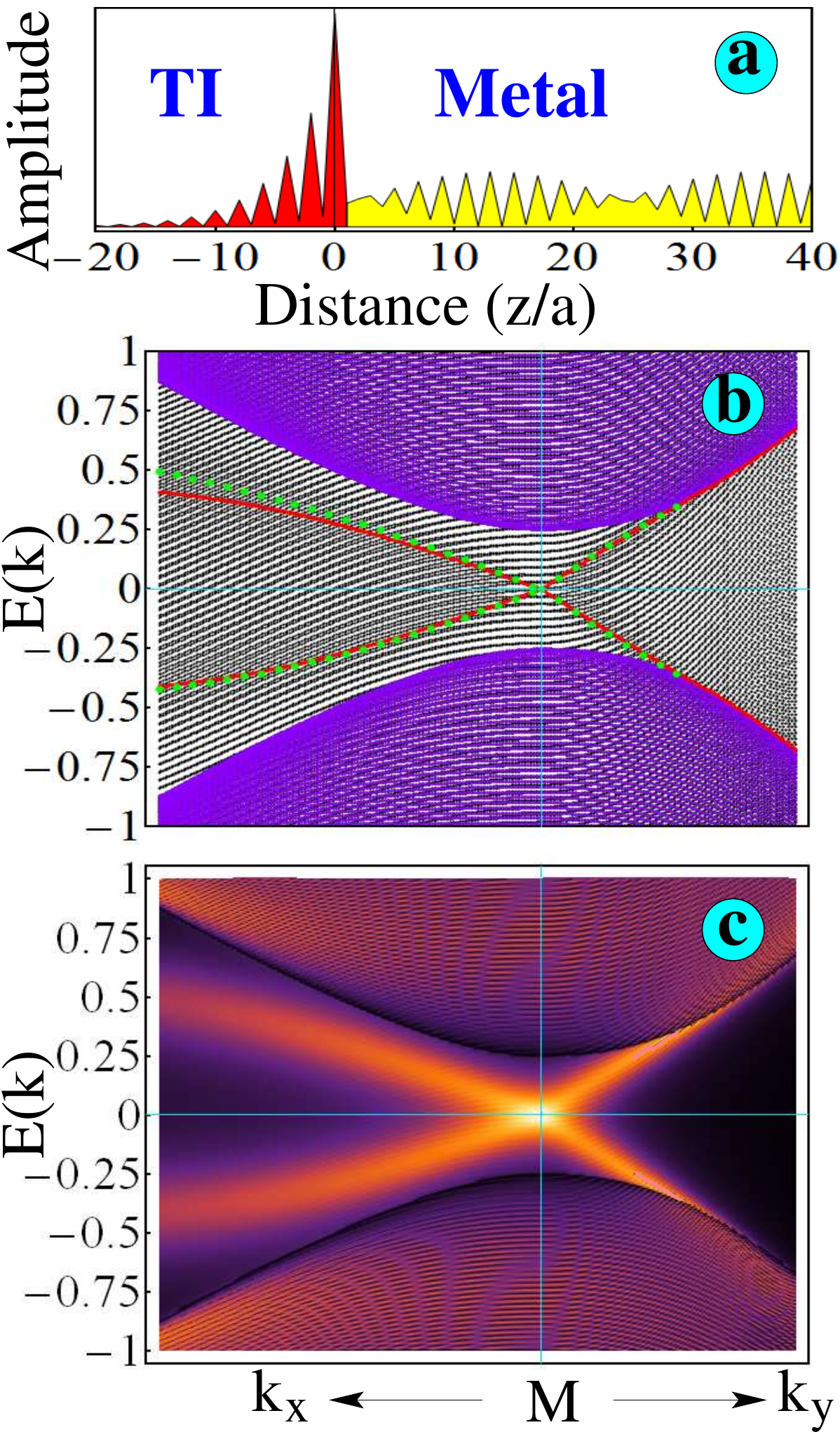}
\vspace{-3mm}
\end{center}
\caption{(Color online) (a) Amplitude of a metallic state hybridized with a topological insulator surface state as function of distance from the interface (in units of interlayer spacing). (b) Spectrum of a topological insulator in contact with a metal (black). The states of a topological insulator with a free surface are also shown (purple - bulk states, red - surface states). (c) Density plot of the total amplitude within a thin region of topological insulator in the vicinity of the interface (red/dark grey area in panel a). Note that, instead of a well-defined surface mode, one has a diffuse distribution of states with boundary contributions. The dispersion of the maxima of this distribution is represented by green points in panel (b).} 
\label{FigContacts}
\end{figure}

For a free surface, the topological insulator spectrum is characterized by a bulk gap occupied in the vicinity of the M-point by an anisotropic Dirac cone associated with gapless surface states. \cite{Stanescu_TI_Proximity_10} These surface states are exponentially localized near the boundary, with a characteristic length scale of a few lattice spacings. When the TI surface is in contact with a metal, the surface states penetrate inside the metal and hybridize with the metallic states. A typical hybridized state is shown in Fig. \ref{FigContacts}a. Note that the amplitude of this state near the boundary of the topological insulator is reduced by a factor $1/L_z^{(m)}$ with respect to the amplitude of a pure surface state with the same energy, where $L_z^{(m)}$ is the width of the metal in the direction perpendicular to the interface. However, the local density of states near the boundary is not reduced, as the number of hybridized states also scales with $L_z^{(m)}$. The spectrum of a topological insulator in contact with a metal is shown in Fig. \ref{FigContacts}b. Also, we calculated for each state the total amplitude within a thin layer of topological insulator in the vicinity of the boundary. The result is shown in Fig. \ref{FigContacts}c. Instead of the sharply defined Dirac cone that characterizes the free surface, one has a diffuse distribution of states with boundary contributions. 

The properties of the interface amplitude distribution shown in Fig. \ref{FigContacts}c, i.e., its width and the dispersion of its maximum, depend on the strength of the coupling between the metal and the topological insulator. If the distribution is sharp enough and the dispersion does not deviate significantly from the Dirac cone, our transport analysis should be applied using bare parameters, i.e., parameters characterizing the spectrum of a free surface. Nonetheless, significant deviations from the free surface dispersion are, in principle, possible. For example, the location of the Dirac point in our model calculation is fixed by symmetry. However, in real systems the energy of the Dirac point can be easily modified by changing the boundary conditions. Consequently, the effective parameters $D_{\rm eff}$ and $A_{\rm eff}$ entering transport coefficients may differ significantly from the corresponding parameters extracted from ARPES measurements. To clarify these issues, as well as to determine the dependence of these effects on the size of the contacts, further studies are necessary. 

\section{Summary}

We have determined the charge conductivity of the surface states of three-dimensional topological insulators basing our theory on the quantum Liouville equation and accounting fully for the absence of backscattering. We identify two contributions to the conductivity with different dependencies on the carrier number density. At low densities the conductivity is $\propto n$ if charged impurity scattering is dominant and is independent of $n$ if short-range scatterers such as surface roughness are dominant. As the density increases an additional term is present in the conductivity, which is $\propto n^{3/2}$ for charged impurity scattering and $\propto n^{1/2}$ for short-range scattering. The ratio of the transport and quantum lifetimes in the region $r_s \ll 4$, which we believe to be the most realistic experimentally, is of the form $b_1 r_s^{-1} - b_2 \ln r_s$ for charged impurity scattering and of the form $b_5 + b_6 r_s \ln r_s$ for surface roughness scattering. The analysis we have presented should enable one to obtain a complete characterization of all the features of topological insulator samples accessible in transport experiments, provided $\varepsilon_F \tau/\hbar \gg 1$ and the chemical potential is below the bottom of the conduction band.

The current form of the theory presented in this article gives results equivalent to those of Boltzmann transport theory. Nevertheless, this theory accounts explicitly for interband coherence effects such as Zitterbewegung, which should become important near the Dirac point in ballistic samples. We expect the advantages of this formulation to become manifest in the presence of a magnetic field. A magnetic field induces additional coherence between the two bands, influences the spin dynamics, and gives rise to a qualitatively different scattering term. The flexibility of the theory outlined in this work enables us to study a variety of multiband effects beyond Boltzmann transport, which we reserve for a future publication.

\acknowledgments

We are grateful to H. D. Drew, N. P. Butch, G. S. Jenkins, J. Paglione, A. B. Sushkov, Roman Lutchyn and Jay Sau for discussions. This work was supported by LPS-NSA-CMTC.

\appendix

\begin{widetext}

\section{Derivations of certain expressions appearing in the scattering term}
\label{sec:app}

Firstly, Eq. (\ref{eq:gsol}) is derived as follows: 
\begin{equation}
\td{g_{{\bm k}{\bm k}'}}{t} + \frac{i}{\hbar} \, [\hat{H}, \hat g]_{{\bm k}{\bm k}'} = - \frac{i}{\hbar} \, [\hat U, \hat f + \hat g]_{{\bm k}{\bm k}'},
\end{equation}
To first order in $\hat{U}$ we do not retain the $\hat{g}$-dependent term on the RHS. We also drop the ${\bm k}$ and ${\bm k}'$ indices for simplicity, writing the equation as an operator equation
\begin{equation}
\td{g}{t} + \frac{i}{\hbar} \, [\hat{H}, \hat g] = - \frac{i}{\hbar} \, [\hat U, \hat f].
\end{equation}
Now let $g = e^{-i\hat{H}t/\hbar} g_I e^{i\hat{H}t/\hbar}$, a transformation equivalent to switching to the interaction picture
\begin{equation}
\arraycolsep 0.3ex
\begin{array}{rl}
\displaystyle \td{g_I}{t} = & \displaystyle - \frac{i}{\hbar} \, e^{i\hat{H}t/\hbar}[\hat U, \hat f]e^{-i\hat{H}t/\hbar} \\ [3ex]
\displaystyle g_I = & \displaystyle - \frac{i}{\hbar} \, \int_{-\infty}^t dt' e^{i\hat{H}t'/\hbar}[\hat U, \hat f(t')]e^{-i\hat{H}t'/\hbar} \\ [3ex]
\displaystyle g = & \displaystyle - \frac{i}{\hbar} \, \int_{-\infty}^t dt' e^{-i\hat{H}(t - t')/\hbar}[\hat U, \hat f(t')]e^{i\hat{H}(t - t')/\hbar} \\ [3ex]
                          = & \displaystyle - \frac{i}{\hbar} \, \int_{0}^{\infty} dt'' e^{-i\hat{H}t''/\hbar}[\hat U, \hat f(t - t'')]e^{i\hat{H}t''/\hbar}
\end{array}
\end{equation}
Projecting onto ${\bm k}$ and ${\bm k}'$ we obtain Eq. (\ref{eq:gsol}). 

We consider in more detail the derivation of the contribution $ \propto ({\bm k} + {\bm k}')$ in the scattering term. The scattering term has the general form (with summation implied over the index $i$)
\begin{equation}
\arraycolsep 0.3ex
\begin{array}{rl}
\displaystyle \hat J(f) = & \displaystyle \frac{n_i}{\hbar^2} \int \frac{d^2k'}{(2\pi)^2} \, |U_{{\bm k}{\bm k}'}|^2 \big(n_{\bm k} -
n_{{\bm k}'} \big) \int^{\infty}_0 dt'\, e^{- \eta t'} e^{- i H_{{\bm k}'} t'} \, e^{i H_{{\bm k}} t'} + h.c. \\ [3ex] 
+ & \displaystyle \frac{n_i}{2\hbar^2} \int \frac{d^2k'}{(2\pi)^2} \, |U_{{\bm k}{\bm k}'}|^2 \big( S_{{\bm k}i} - S_{{\bm k}'i}\big)
\int^{\infty}_0 dt'\, e^{-\eta t'} e^{- i H_{{\bm k}'} t'} \sigma_i  \, e^{i H_{{\bm k}} t'} + h.c. 
\end{array}
\end{equation}
In the time evolution operator one substitutes the general expression for the band Hamiltonian $H_{0{\bm k}} = Dk^2 + (1/2) \, {\bm \sigma} \cdot {\bm \Omega}_{\bm k}$, where ${\bm \Omega}_{\bm k} = 2 A {\bm k}$. In what follows we shall shorten the time evolution operators by omitting the factors of $\hbar$. These will be restored in the final expressions. The time evolution operator can be written as
\begin{equation}
\arraycolsep 0.3ex
\begin{array}{rl}
\displaystyle e^{- i H_{{\bm k}} t} = & \displaystyle e^{ -i Dk^2 t}\bigg( \cos\frac{\Omega_{{\bm k}} t}{2} - i\, {\bm \sigma}\cdot\hat{\bm \Omega}_{{\bm k}}\, \sin\frac{\Omega_{{\bm k}} t}{2} \bigg).
\end{array}
\end{equation}
We recall the identity
\begin{equation}
({\bm \sigma}\cdot\hat{\bm \Omega}_{{\bm k}}) \, ({\bm \sigma}\cdot\hat{\bm \Omega}_{{\bm k}'}) = \hat{\bm \Omega}_{{\bm k}}\cdot\hat{\bm \Omega}_{{\bm k}'} + i \, {\bm \sigma}\cdot\hat{\bm \Omega}_{{\bm k}} \times \hat{\bm \Omega}_{{\bm k}'}.
\end{equation}
This allows us to express the products of two time evolution operators as
\begin{equation}
\arraycolsep 0.3ex
\begin{array}{rl}
\displaystyle e^{- i H_{{\bm k}} t} e^{ i H_{{\bm k}'}t}
= & \displaystyle e^{ i(Dk^{'2} - Dk^2) t} [\cos\frac{\Omega_{{\bm k}'}
t}{2}\cos\frac{\Omega_{{\bm k}} t}{2} - i\, {\bm
\sigma}\cdot\hat{\bm \Omega}_{{\bm k}}\,\cos\frac{\Omega_{{\bm
k}'} t}{2} \sin\frac{\Omega_{{\bm k}} t}{2} + i\, {\bm
\sigma}\cdot\hat{\bm \Omega}_{{\bm k}'}\, \sin\frac{\Omega_{{\bm
k}'} t}{2}\cos\frac{\Omega_{{\bm k}} t}{2}
\\ [3ex] & \displaystyle
+ (\hat{\bm \Omega}_{{\bm k}'}\cdot\hat{\bm \Omega}_{{\bm k}} + i
{\bm \sigma}\cdot\hat{\bm \Omega}_{{\bm k}}\times\hat{\bm
\Omega}_{{\bm k}'})\, \sin\frac{\Omega_{{\bm k}'} t}{2}\,
\sin\frac{\Omega_{{\bm k}} t}{2}] \\ [3ex]
\displaystyle e^{- i H_{{\bm k}} t} {\bm \sigma} \, e^{ i
H_{{\bm k}'}t} = & \displaystyle e^{ i(Dk^{'2} - Dk^2) t} [{\bm \sigma}
\cos\frac{\Omega_{{\bm k}'} t}{2}\cos\frac{\Omega_{{\bm k}} t}{2}
- i\, ({\bm \sigma}\cdot\hat{\bm \Omega}_{{\bm k}}) \, {\bm
\sigma} \, \cos\frac{\Omega_{{\bm k}'} t}{2}
\sin\frac{\Omega_{{\bm k}} t}{2} + i\, {\bm \sigma} \, ({\bm
\sigma}\cdot\hat{\bm \Omega}_{{\bm k}'}) \, \sin\frac{\Omega_{{\bm
k}'} t}{2}\cos\frac{\Omega_{{\bm k}} t}{2}
\\ [3ex] & \displaystyle + ({\bm \sigma}\cdot\hat{\bm \Omega}_{{\bm
k}}) \, {\bm \sigma} \, ({\bm \sigma}\cdot\hat{\bm \Omega}_{{\bm
k}'}) \, \sin\frac{\Omega_{{\bm k}'} t}{2}\,
\sin\frac{\Omega_{{\bm k}} t}{2}].
\end{array}
\end{equation}

The time integrals of these products of exponentials and trigonometric functions give
\begin{equation}
\arraycolsep 0.3ex
\begin{array}{rl}
\displaystyle e^{ i(Dk^{'2} - Dk^2) t} \cos\frac{\Omega_{{\bm k}} t}{2}\cos\frac{\Omega'_{{\bm
k}} t}{2} \rightarrow & \displaystyle \frac{\pi\hbar}{4}\,
[\delta(\epsilon_+ - \epsilon'_+) + \delta(\epsilon_- -
\epsilon'_-) + \delta(\epsilon_+ - \epsilon'_-) +
\delta(\epsilon_- - \epsilon'_+)]
\\ [3ex]
\displaystyle e^{ i(Dk^{'2} - Dk^2) t}  \cos\frac{\Omega_{{\bm k}'} t}{2} \sin\frac{\Omega_{{\bm
k}} t}{2} \rightarrow & \displaystyle \frac{\pi\hbar}{4i}\,
[\delta(\epsilon'_+ - \epsilon_-) + \delta(\epsilon'_- -
\epsilon'_-) - \delta(\epsilon'_+ - \epsilon_+) -
\delta(\epsilon'_- - \epsilon_+)]
\\ [3ex]
\displaystyle e^{ i(Dk^{'2} - Dk^2) t}  \sin\frac{\Omega_{{\bm k}'} t}{2}\cos\frac{\Omega_{{\bm
k}} t}{2} \rightarrow & \displaystyle \frac{\pi\hbar}{4i}\,
[\delta(\epsilon'_+ - \epsilon_+) + \delta(\epsilon'_+ -
\epsilon_-) - \delta(\epsilon'_- - \epsilon_-) -
\delta(\epsilon'_- - \epsilon_+)]
\\ [3ex]
\displaystyle e^{ i(Dk^{'2} - Dk^2) t}  \sin\frac{\Omega_{{\bm k}} t}{2}\, \sin\frac{\Omega'_{{\bm
k}} t}{2} \rightarrow & \displaystyle - \frac{\pi\hbar}{4}\,
[\delta(\epsilon_+ - \epsilon'_-) + \delta(\epsilon_- -
\epsilon'_+) - \delta(\epsilon_- - \epsilon'_-) -
\delta(\epsilon_+ - \epsilon'_+)]
\end{array}
\end{equation}
The Hermitian conjugates are trivial. For the scalar part of the density matrix, letting $\hat{\bm \Omega}_{{\bm k}} = \hat{\bm k}$ the above results in
\begin{equation}
\arraycolsep 0.3ex
\begin{array}{rl}
\displaystyle e^{- i\hat H_{{\bm k}'} t'} e^{ i \hat H_{{\bm k}}t'} + e^{ - i \hat H_{{\bm k}}t'} e^{ i \hat H_{{\bm k}'}t'} \rightarrow & \displaystyle \frac{\pi\hbar}{2}\, (1 + \hat{\bm k}\cdot\hat{\bm k}') \, [\delta(\epsilon'_+ - \epsilon_+) + \delta(\epsilon'_- - \epsilon_-)] \\ [3ex]
+ & \displaystyle \frac{\pi\hbar}{2}\, {\bm \sigma}\cdot(\hat{\bm k} + \hat{\bm k}') \, [\delta(\epsilon'_+ - \epsilon_+) - \delta(\epsilon'_- - \epsilon_-)].
\end{array}
\end{equation}
For the spin-dependent part of the density matrix,
\begin{equation}
\arraycolsep 0.3ex
\begin{array}{rl}
\displaystyle e^{- i H_{{\bm k}} t} {\bm \sigma} \, e^{ i H_{{\bm k}'}t} \rightarrow & \displaystyle \frac{\pi\hbar}{4} \, [{\bm \sigma} + ({\bm \sigma}\cdot\hat{\bm k}) \, {\bm \sigma} \, ({\bm \sigma}\cdot\hat{\bm k}')] \, [\delta(\epsilon'_+ - \epsilon_+) + \delta(\epsilon'_- - \epsilon_-)]\\ [3ex]

+ & \displaystyle \frac{\pi\hbar}{4} \, [({\bm \sigma}\cdot\hat{\bm k}) \, {\bm \sigma} + {\bm \sigma} \, ({\bm \sigma}\cdot\hat{\bm k}')] \, [\delta(\epsilon'_+ - \epsilon_+) - \delta(\epsilon'_- - \epsilon_-)] \\ [3ex]

\displaystyle e^{- i H_{{\bm k}'} t'}{\bm \sigma}\, e^{ i H_{{\bm k}}t'} + e^{ - i \hat H_{{\bm k}}t'} {\bm \sigma}\, e^{ i \hat H_{{\bm k}'}t'} \rightarrow & \displaystyle \frac{\pi\hbar}{4} \, \{ [2{\bm \sigma} + ({\bm \sigma}\cdot\hat{\bm k}) \, {\bm \sigma} \, ({\bm \sigma}\cdot\hat{\bm k}') + ({\bm \sigma}\cdot\hat{\bm k}') \, {\bm \sigma} \, ({\bm \sigma}\cdot\hat{\bm k}) ] \, [\delta(\epsilon'_+ - \epsilon_+) + \delta(\epsilon'_- - \epsilon_-)] \\ [3ex]

& \displaystyle + \frac{\pi\hbar}{4} \, [({\bm \sigma}\cdot\hat{\bm k}) \, {\bm \sigma} + {\bm \sigma} \, ({\bm \sigma}\cdot\hat{\bm k}') + ({\bm \sigma}\cdot\hat{\bm k}') \, {\bm \sigma} + {\bm \sigma} \, ({\bm \sigma}\cdot\hat{\bm k})] \, [\delta(\epsilon'_+ - \epsilon_+) - \delta(\epsilon'_- - \epsilon_-)] \}.
\end{array}
\end{equation}
The product of three sigma matrices simplifies as follows
\begin{equation}
\arraycolsep 0.3ex
\begin{array}{rl}
\displaystyle ({\bm \sigma}\cdot\hat{\bm k}) \, {\bm \sigma} \, ({\bm \sigma}\cdot\hat{\bm k}') + ({\bm \sigma}\cdot\hat{\bm k}') \,
{\bm \sigma} \, ({\bm \sigma}\cdot\hat{\bm k}) = & \displaystyle 2(\hat{\bm k} \cdot{\bm \sigma}) \, \hat{\bm k}' -
2(\hat{\bm k} \cdot \hat{\bm k}')\, {\bm \sigma} + 2\hat{\bm k} \, (\hat{\bm k}'\cdot{\bm \sigma}),
\end{array}
\end{equation}
yielding finally
\begin{equation}
\arraycolsep 0.3ex
\begin{array}{rl}
\displaystyle e^{- i H_{{\bm k}'} t'}{\bm \sigma}\, e^{ i H_{{\bm k}}t'} + e^{ - i H_{{\bm k}}t'} {\bm \sigma}\, e^{ i H_{{\bm k}'}t'} \rightarrow & \displaystyle \frac{\pi\hbar}{2} \, [{\bm \sigma}(1 - \hat{\bm k} \cdot \hat{\bm k}') + (\hat{\bm k} \cdot{\bm \sigma}) \, \hat{\bm k}' + \hat{\bm k} \, (\hat{\bm k}'\cdot{\bm \sigma})] \, [\delta(\epsilon'_+ - \epsilon_+) + \delta(\epsilon'_- - \epsilon_-)]\\ [3ex]
+ & \displaystyle \frac{\pi\hbar}{2} \, (\hat{\bm k} + \hat{\bm k}') \, [\delta(\epsilon'_+ - \epsilon_+) - \delta(\epsilon'_- - \epsilon_-)].
\end{array}
\end{equation}
The scattering term takes the form $\hat J(f) = \hat J(n) + \hat J(S)$
\begin{equation}
\arraycolsep 0.3ex
\begin{array}{rl}
\displaystyle \hat J(n) = & \displaystyle \frac{\pi n_i}{2\hbar} \int \frac{d^2k'}{(2\pi)^2} \, |\bar{U}_{{\bm k}{\bm k}'}|^2 \big(n_{\bm k} -
n_{{\bm k}'} \big) (1 + \hat{\bm k}\cdot\hat{\bm k}') \, [\delta(\epsilon'_+ - \epsilon_+) + \delta(\epsilon'_- - \epsilon_-) ] \\ [3ex]
+ & \displaystyle \frac{\pi n_i}{2\hbar} \int \frac{d^2k'}{(2\pi)^2} \, |\bar{U}_{{\bm k}{\bm k}'}|^2 \big(n_{\bm k} -
n_{{\bm k}'} \big)  {\bm \sigma}\cdot(\hat{\bm k} + \hat{\bm k}')\, [\delta(\epsilon'_+ - \epsilon_+) - \delta(\epsilon'_- - \epsilon_-) ] \\ [3ex]
\displaystyle \hat J(S) = & \displaystyle \frac{\pi n_i}{4\hbar} \int \frac{d^2k'}{(2\pi)^2} \, |\bar{U}_{{\bm k}{\bm k}'}|^2 \big( {\bm S}_{{\bm k}} - {\bm S}_{{\bm k}'}\big)\cdot [{\bm \sigma}(1 - \hat{\bm k} \cdot \hat{\bm k}') + (\hat{\bm k} \cdot{\bm \sigma}) \, \hat{\bm k}' + \hat{\bm k} \, (\hat{\bm k}'\cdot{\bm \sigma}) ] \, [\delta(\epsilon'_+ - \epsilon_+) + \delta(\epsilon'_- - \epsilon_-) ] \\ [3ex]
+ & \displaystyle \frac{\pi n_i}{4\hbar} \int \frac{d^2k'}{(2\pi)^2} \, |\bar{U}_{{\bm k}{\bm k}'}|^2 \big( {\bm S}_{{\bm k}} - {\bm S}_{{\bm k}'}\big)\cdot (\hat{\bm k} + \hat{\bm k}') \,[\delta(\epsilon'_+ - \epsilon_+) - \delta(\epsilon'_- - \epsilon_-) ].
\end{array}
\end{equation}
Note the factor of 2 difference between $\hat{J}(n)$ and $\hat{J}(S)$, which arises solely from our definition $H_{0{\bm k}} = Dk^2 + (1/2) \, {\bm \sigma} \cdot {\bm \Omega}_{\bm k}$. Note also that there are no second-order terms in the Pauli matrices $\sigma_i$. From the algebra of these matrices we know that 
\begin{equation}
\begin{array}{rl}
\displaystyle \sigma_i \sigma_j = & \displaystyle \delta_{ij} + \epsilon_{ijk}  \sigma_k \\ [3ex]
\displaystyle [\sigma_i, \sigma_j] = & \displaystyle 2 \epsilon_{ijk}  \sigma_k \\ [3ex]
\displaystyle \{ \sigma_i, \sigma_j \} = & \displaystyle 2\delta_{ij}.
\end{array}
\end{equation}
in other words the second-order terms can give zero, the identity matrix [leading to the terms $\propto \big( {\bm S}_{{\bm k}} - {\bm S}_{{\bm k}'}\big)\cdot (\hat{\bm k} + \hat{\bm k}')$], or one Pauli matrix.

\end{widetext}


\end{document}